\documentclass[12pt,onecolumn,draftclsnofoot]{IEEEtran}

\ifCLASSINFOpdf  
\else 
\fi

\usepackage[colorlinks,urlcolor=blue,linkcolor=blue,citecolor=blue]{hyperref}

\usepackage{color,array,amsthm}
\usepackage{algpseudocode}
\usepackage{graphicx}
\usepackage{subfigure}
\usepackage{bm}
\usepackage{textcomp}
\usepackage{amsmath}
\usepackage{lipsum}
\usepackage{cuted}\stripsep -3pt plus 3pt minus 2pt
\usepackage{amsthm}
\usepackage{amssymb}
\usepackage{amsfonts}
\usepackage{latexsym}
\usepackage{amsfonts}
\usepackage{algorithm}

\usepackage{booktabs}
\usepackage{threeparttable}
\usepackage{multirow}
\usepackage{algorithmicx}  
\usepackage{algpseudocode}

\graphicspath{ {figures/} }

\begin{document}
%
\title{Underdetermined  2D-DOD and 2D-DOA  Estimation for Bistatic Coprime EMVS-MIMO Radar: From the Difference Coarray Perspective}
%
%
%

\author{Qianpeng XIE,
        Yihang DU, 
        He WANG,
        Xiaoyi PAN, 
        and~Feng ZHAO
\thanks{This work was supported in part by the National Natural Science Foundation of China under Grant 61890545, 61890542 and 61890540. Qianpeng XIE, Xiaoyi PAN, Feng ZHAO are with the State Key Laboratory of Complex Electromagnetic Environment Effects on Electronics and Information System, National University of Defense Technology, Changsha 410073, China(email:13721038905@163.com, mrpanxy@nudt.edu.cn, zhfbee@tom.com). Yihang DU is with the Sixty-Third Research Institute, National University of Defense Technology, Nanjing 210007, China(email:dyhcsl1991@163.com). He WANG is with the School of Electronic Science, National University of Defense Technology, Changsha 410073, China.(email:hw8g16@soton.ac.uk)(Corresponding author: Qianpeng XIE)}
\thanks{Manuscript received April 19, 2005; revised August 26, 2015.}}

\markboth{Journal of \LaTeX\ Class Files,~Vol.~14, No.~8, August~2015}%
{Shell \MakeLowercase{\textit{et al.}}: Bare Demo of IEEEtran.cls for IEEE Journals}

\maketitle

\begin{abstract}
In this paper, the underdetermined 2D-DOD and 2D-DOA estimation for bistatic coprime EMVS-MIMO radar is considered. Firstly, a 5-D tensor model {was} constructed by using the {multi-dimensional} space-time characteristics of the {received} data. Then, an 8-D tensor {has been} obtained by using the auto-correlation calculation. To obtain the difference coarrays of transmit and receive EMVS,  the de-coupling process between the spatial response of EMVS and the steering vector is inevitable. Thus, a new  6-D tensor can be constructed via the tensor permutation and the generalized tensorization of the canonical polyadic decomposition. {According} to the theory of the Tensor-Matrix Product operation, the {duplicated elements}  in the difference coarrays can be removed by {the utilization of} two designed selection matrices. 
Due to the {centrosymmetric} geometry of the difference coarrays, two DFT beamspace matrices {were subsequently} designed to convert the complex steering matrices into the real-valued {ones, whose advantage is to} improve the estimation accuracy of the 2D-DODs and 2D-DOAs. {Afterwards}, a {third-order} tensor with the third-way fixed at 36 {was} constructed  {and} the Parallel Factor algorithm {was deployed}, which can yield the closed-form automatically paired 2D-DOD and 2D-DOA estimation. The simulation results show that  the proposed algorithm can exhibit superior estimation performance for the underdetermined 2D-DOD and 2D-DOA estimation.
\end{abstract}


%
\IEEEpeerreviewmaketitle

\section{Introduction}

\IEEEPARstart{S}{tudies} about the electromagnetic vector sensors (EMVS) have attracted extensive attention due to {their} excellent measurement capabilities of angle parameters and polarization parameters  \cite{WongK}-\cite{LanX}. Compared with the scalar array, EMVS with three orthogonal electric dipoles and three orthogonal magnetic dipoles can obtain the {electromagnetic} field vector information of the targets. Recently, in order to measure the polarization information of targets in bistatic MIMO radars, the EMVS {has been} used as the transmit and receive array  in the bistatic MIMO radar system.
\par {For the achievement of} multi-dimensional parameter estimation in bistatic EMVS-MIMO radar, the ESPRIT algorithm {was} firstly proposed in \cite{Chintagunta}. But, this algorithm has high computational complexity due to the singular value decomposition on the covariance matrix of the array received data. In \cite{LiuT}, a computationally fridendly PM estimator {was} proposed to construct the virtual direction matrix for bistatic EMVS-MIMO radar. In \cite{ MaoC}, a tensor subspace based algorithm {has been} developed to improve the estimated performance by exploiting the inherent {multi-dimensional} structure of the received data. Additional optimization process in \cite{Chintagunta}-\cite{MaoC} is inevitable {because of} the pair matching between 2D-DOD and 2D-DOA. Thus, an improved PM algorithm in \cite{WenF} {has been} {raised} to realize the {automatic pair of} 2D-DOD and 2D-DOA{, and this method}  can further decrease the computational complexity. The disadvantage of the PM algorithms in \cite{LiuT}  and \cite{WenF}  is the relatively poor estimation performance. In \cite{WenFShi}, the PARAFAC algorithm {was utilized} to realize the automatically paired  2D-DOD and 2D-DOA estimation. In \cite{XianpengWang}, the bistatic coprime EMVS-MIMO radar {was} designed to {enhance} the 2D-DOD and 2D-DOA estimaiton performance. 
\par {It is apparent that the above-mentioned algorithms do not utilize the inherent difference coarray of transmit and receive EMVS, when conducting multi-dimensional parameter estimation in bistatic EMVS MIMO radar.} According to 
\cite{VaidyanathanPP}-\cite{ZhengCheng},  the difference coarray of the original array can offer larger array apertures, which means that the underdetermined DOA estimation {is obtainable.} For bistatic EMVS-MIMO radar, the {challenge}  for constructing the difference coarray of the transmit  and receive EMVS is the coupling in the spatial response {alongside} the steering vector. In this paper, the de-coupling process {has been} completed with the help of tensor operation. {To} estimate more targets than elements, the difference coarray of transmit and receive EMVS {has been attained} by the high order tensor permutation {along with} the generalized tensoriation of the canonical polyadic decomposition. {In addition}, two designed selection matrices {have been} used to remove the repeated elements in the difference coarray via the Tensor-Matrix product rule. The geometry of the difference coarray {for both transmit and receive EMVS } exhibit {centrosymmetry}. 
{As discussed in}  \cite{DongheLiu}-\cite{Sheng}, the DFT beamspace matrix can be {exploited} to convert the complex array manifold matrix with  centro-symmetric structure into the real-value one. 
This process can {enhance} the 2D-DOD and 2D-DOA estimation performance in the bistatic EMVS-MIMO radar. {Besides}, a  {third-order} tensor {has been} constructed with the dimension of the third-way fixed at 36. Finally, The {automatically}  paired transmit/receive elevation angle, transmit/receive azimuth angle, transmit/receive polarization angle and transmit/receive polarization phase difference can be accurately derived from the estimated transmit and receive factor {matrices} {acquired}  by the PARAFAC algorithm. Simulation results {have verified its}  {effectiveness}.
\par The paper is organized as follows.  Section II describes the nomenclature {employed}  in this {publication}. Section III formulates the signal model of the bistatic coprime EMVS-MIMO radar. Section IV reports the detailed derivation process of the proposed algorithm. Section V presents {simulations} to verify the {effectiveness} of the proposed algorithm. Section VI {provides a summary}.

\section{Nomenclature}
\begin{itemize}
	\item $\boldsymbol{A}^{*}$:  The complex conjugate  of $\boldsymbol{A}$
	\item $\boldsymbol{A}^{T}$:  The transpose of $\boldsymbol{A}$
	\item $\boldsymbol{A}^{H}$: The Hermitian transpose of $\boldsymbol{A}$ 
	\item $\boldsymbol{A}^{-1}$: The inverse of $\boldsymbol{A}$ 
	\item $\boldsymbol{A}^{\dagger}$: The pseudo-inverse of $\boldsymbol{A}$ 
	\item $\boldsymbol{a}\otimes\boldsymbol{b}$: The Kronecker product of $\boldsymbol{a}$ and $\boldsymbol{b}$
	\item $\boldsymbol{A}\odot\boldsymbol{B}$: The Khatri-Rao product of $\boldsymbol{A}$ and $\boldsymbol{B}$
	\item $\boldsymbol{A}\oplus\boldsymbol{B}$: The Hadamard product of $\boldsymbol{A}$ and $\boldsymbol{B}$
	\item   $\boldsymbol{a}\circ\boldsymbol{b}$: The vector outer product of $\boldsymbol{a}$ and $\boldsymbol{b}$
	\item $\boldsymbol{a}\times\boldsymbol{b}$ : The vector-cross-product of $\boldsymbol{a}$ and $\boldsymbol{b}$
	\item $(\boldsymbol{\Pi})^{\perp}_{\boldsymbol{A}}$ : The  projection matrix of the matrix $\boldsymbol{A}$
	\item $sin(\theta)$ : The sine of $\theta$
	\item $cos(\theta)$:  The cosine of $\theta$
	\item $tan(\theta)$: The tangent of $\theta$
	\item $arcsin(\theta)$: The  inverse sine of $\theta$  
	\item $arccos(\theta)$:  The inverse cosine of $\theta$  
	\item $arctan(\theta)$: The inverse tangent of $\theta$  
	\item $\angle(\boldsymbol{a})$:  The phase angles of $\boldsymbol{a}$ 
	\item $real(\boldsymbol{A})$:  The real part of $\boldsymbol{A}$ 
	\item $min[a,b]$: {smaller} value taken from $a$ or $b$.
	\item $\mathcal{D}_{i}(\boldsymbol{A} )$: The diagonal matrix with the  $i$-th colume\\ \qquad \qquad elements  from $\boldsymbol{A} $.
	\item $\boldsymbol{I}_{N} $: The $N\times N$ identity matrix 
	\item $\boldsymbol{0}_{N} $: The $N\times N$ matrix of zeros.
	\item $\boldsymbol{1}_{N} $:  The $N\times N$ matrix of ones. 
	\item $\kappa_{{\boldsymbol{A}}} $:  The $\kappa$-rank of ${\boldsymbol{A}}$. 
	\item $\textbf{Some Definations of Coprime Array in}$ {\cite{VaidyanathanPP}-\cite{LiuC}}\\	
	$\textbf{Definition 1}$ ($\mathbb{D}$, Difference Coarray). For a sparse coprime array specified by an integer set $\mathbb{S}$, its difference coarray $\mathbb{D}$ is defined as 
	\begin{equation}
	\begin{aligned}
	\mathbb{D}=\lbrace n_{1}-n_{2}|n_{1},n_{2}\in  \mathbb{S} \rbrace
	\end{aligned}   \label{1}
	\end{equation}\\
	$\textbf{Definition 2}$ ($\mathbb{U}$, maximum central contiguous ULA ). Let $\mathbb{S}$ denote a coprime array and $\mathbb{D}$ be its difference coarray. The maximum central contiguous ULA segment in $\mathbb{D}$ is
	\begin{equation}
	\begin{aligned}
	\mathbb{U}=\lbrace m|-|m|,\cdots,-1,0,1,\cdots,|m|\rbrace
	\end{aligned}   \label{2} 
	\end{equation}\\
	$\textbf{Definition 3}$ ($\mathbb{W}$, Weight Functions). The weight function $w(m)$ of an array $\mathbb{S}$ is defined as the number of sensor pairs that lead to coarray index $m$, i.e.,$|\lbrace (n_{1},n_{2} \in \mathbb{S}^{2}) | n_{1}-n_{2}=m\rbrace  |$
	\item  $\textbf{Some Definations of Tensor in} $  {\cite{GKolda}-\cite{WRao}} \\ 
	$\textbf{Definition 4}$ (Mode-$n$ Tensor-Matrix Product). The mode-$n$ product of an $N$-order tensor  set $\mathcal{X}\in \mathbb{C}^{I_{1}\times I_{2}\times \cdots \times I_{N} }$ and a matrix ${\boldsymbol{A}}\in \mathbb{C}^{J_{n}\times I_{n} }$ is defined as
	\begin{equation}
	\begin{aligned}
	\mathcal{Y}&= \mathcal{X}\times_{n}{\boldsymbol{A}}\in \mathbb{C}^{I_{1}\times\cdots \times I_{n-1}\times J_{n}\times I_{n+1} \times\cdots \times I_{N} }
	\end{aligned}    \label{3}
	\end{equation}\\
	$\textbf{Definition 5}$ (Outer product of two tensors). The outer product of two tensors $\mathcal{A}\in \mathbb{C}^{I_{1}\times I_{2}\times \cdots \times I_{N} }$ and $\mathcal{B}\in \mathbb{C}^{J_{1}\times J_{2}\times \cdots \times J_{M} }$ is defined as
	\begin{equation}
	\begin{aligned}
	\mathcal{Y}&= \mathcal{A}\circ \mathcal{B}\in \mathbb{C}^{I_{1}\times\cdots \times I_{N}\times J_{1}\times J_{2} \times\cdots \times J_{M} }
	\end{aligned}   \label{4} 
	\end{equation}\\
	whose elements are ${y}_{i_{1},i_{2},\cdots,i_{N}j_{1},j_{2},\cdots,j_{M} }=a_{i_{1},i_{2},\cdots,i_{N}}\cdot b_{j_{1},j_{2},\cdots,j_{M} }$
	\\
	$\textbf{Definition 6}$ (PARAFAC decomposition). For an $N$th-order tensor $\mathcal{X}\in \mathbb{C}^{I_{1}\times I_{2}\times \cdots \times I_{N} }$, it can be expressed as a linear combination of rank-1 tensors as follows:	
	\begin{equation}
	\begin{aligned}
	\mathcal{X}= \sum_{k=1}^{K}{\boldsymbol{{a}}_{k1}}\circ {\boldsymbol{{a}}_{k2}}\circ {\cdots}\circ {\boldsymbol{{a}}_{kN}} 
	\end{aligned}    \label{5}
	\end{equation}\\
	where ${\boldsymbol{{a}}_{kn}}\in \mathbb{C}^{I_{n}\times 1},k=1,2,\cdots,K$ denotes the rank-1 tensor. \\
	$\textbf{Definition 7}$ (Generalized Tensorization of a PARAFAC model). For an $N$th-order PARAFAC model  $\mathcal{A}=\sum_{k=1}^{K}{\boldsymbol{{a}}_{1k}}\circ {\boldsymbol{{a}}_{2k}}\circ {\cdots}\circ {\boldsymbol{{a}}_{Nk}} \in \mathbb{C}^{I_{1}\times I_{2}\times \cdots \times I_{N} }$ with ${\boldsymbol{{a}}_{nk}} \in \mathbb{C}^{I_{n}}(n=1,\cdots,N)$, let the ordered sets $\mathbb{A}_{j}=[r_{j,1},r_{j,2},\cdots,r_{j,L_{j}}]$ for $j=1,2,\cdots,J$ be a partitioning of the dimensions 
	$\mathbb{A}=[1,2,\cdots,N]$, the generalized tensorization of $\mathcal{A}$ is defined as	
	\begin{equation}
	\begin{aligned}
	\mathcal{A}_{\mathbb{A}_{1},\mathbb{A}_{2},\cdots,\mathbb{A}_{J}}&= \sum_{k=1}^{K}{\boldsymbol{{b}}_{1k}}\circ {\boldsymbol{{b}}_{2k}}\circ {\cdots}\circ {\boldsymbol{{b}}_{Jk}} \\
	& \in \mathbb{C}^{I_{R_{1}}\times I_{R_{2}}\times \cdots \times I_{R_{J}} } 
	\end{aligned}    \label{6}
	\end{equation}\\
	where $I_{R_{j}}={\Pi}_{l=1}^{L_{j}}I_{r_{j,l}}$ and ${\boldsymbol{{b}}_{jk}}= \boldsymbol{{a}}_{r_{j},L_{j}}\otimes \boldsymbol{{a}}_{r_{j},L_{j}-1}$ $\otimes {\cdots}\otimes {\boldsymbol{{a}}_{r_{j},1}} $.
	\\
	
\end{itemize}
\section{DATA MODEL}
\begin{figure}[htbp]
	\centerline{\includegraphics[width=28.5pc]{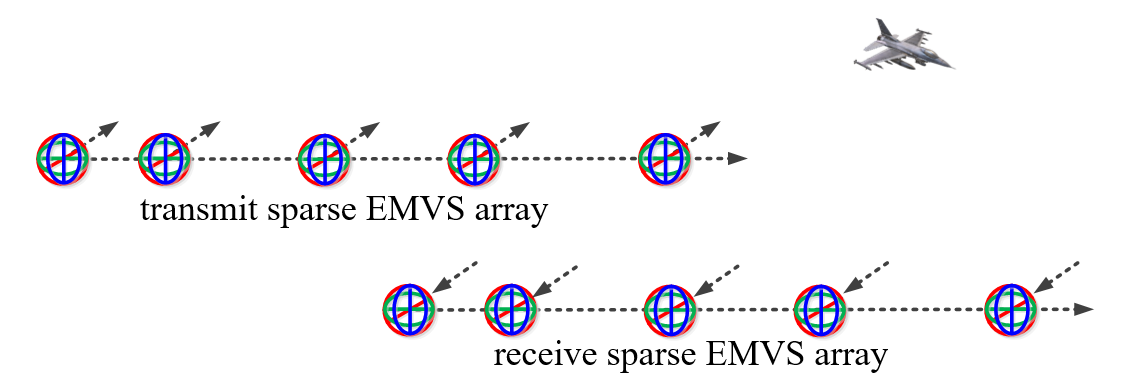}}
	\caption{Bistatic coprime EMVS-MIMO radar system. }
	\label{array1}
\end{figure}

{A bistatic coprime EMVS-MIMO radar system with $M$ transmit EMVS and $N$ receive EMVS is demonstrated in Fig. \ref{array1}. }  Let the array positions of transmit EMVS and receive EMVS are $\boldsymbol{L}_{t}$ and $\boldsymbol{L}_{r}$, respectively
\begin{equation}
\begin{aligned}
\boldsymbol{L}_{t}&= \lbrace M_{1}m_{2}|0\leq m_{2}\leq M_{2}-1\rbrace
\cup  \lbrace M_{2}m_{1}|0\leq m_{1}\leq 2M_{1}-1\rbrace
\end{aligned}  \label{7}  
\end{equation}
\begin{equation}
\begin{aligned}
\boldsymbol{L}_{r}&= \lbrace N_{1}n_{2}|0\leq n_{2}\leq N_{2}-1\rbrace\cup  \lbrace N_{2}n_{1}|0\leq n_{1}\leq 2N_{1}-1\rbrace
\end{aligned}   \label{8} 
\end{equation}
\par Assuming the number of targets is $K$, the transmit steering vector and receive steering vector of the bistatic coprime EMVS-MIMO radar can be expressed as
\begin{equation}
\begin{aligned}
\boldsymbol{c}_{t_{k}} = \boldsymbol{a}_{t_{k}}\otimes \boldsymbol{q}_{t_{k}}(\theta_{t_{k}},\phi_{t_{k}},\gamma_{t_{k}},\eta_{t_{k}})
\end{aligned}  \label{9} 
\end{equation}
\begin{equation}
\begin{aligned}
\boldsymbol{c}_{r_{k}} = \boldsymbol{a}_{r_{k}}\otimes \boldsymbol{q}_{r_{k}}(\theta_{r_{k}},\phi_{r_{k}},\gamma_{r_{k}},\eta_{r_{k}})
\end{aligned}   \label{10}
\end{equation}
where $\theta_{t_{k}},\theta_{r_{k}}\in [0,\pi )$, $\phi_{t_{k}},\phi_{r_{k}}\in [0,2\pi )$, $\gamma_{t_{k}},\gamma_{r_{k}}\in [0,\pi/2 )$ and $\eta_{t_{k}},\eta_{r_{k}}\in [-\pi,\pi )$  denote the transmit/receive elevation angle, transmit/receive azimuth angle, transmit/receive polarization angle and transmit/receive polarization phase difference, respectively.  $\boldsymbol{q}_{t_{k}}(\theta_{t_{k}},\phi_{t_{k}},\gamma_{t_{k}},\eta_{t_{k}})$ and $\boldsymbol{q}_{r_{k}}(\theta_{r_{k}},\phi_{r_{k}},\gamma_{r_{k}},\eta_{r_{k}}) $ represent the spatial response of the transmit EMVS and the receive EMVS of the $k$-th target{, respectively}. ${\boldsymbol{a}_{t_{k}}=}{[
	e^{-j2\pi d_{t0} \sin(\theta_{t_{k}})},\cdots,e^{-j2\pi d_{tM} \sin(\theta_{t_{k}})}
	]^{T}}$,  ${\boldsymbol{a}_{r_{k}}=}{[
	e^{-j2\pi d_{r0} \sin(\theta_{r_{k}})}, }$ ${  \cdots, e^{-j2\pi d_{rN} \sin(\theta_{r_{k}})}
	]^{T}}$.  For an EMVS, the spatial response $\boldsymbol{q}$ can be expressed as
\begin{equation}
\begin{aligned}
&\boldsymbol{q}(\theta,\phi,\gamma,\eta)        
= \\
&\underbrace{\begin{bmatrix}
	\cos(\phi)\cos(\theta) & -\sin(\phi)\\
	\sin(\phi)\cos(\theta)  &\cos(\phi)  \\
	-\sin(\theta)  & 0 \\
	- \sin(\phi)   & -\cos(\phi)\cos(\theta)\\
	\cos(\phi) &   -\sin(\phi)\cos(\theta) \\
	0 &    \sin(\theta)\\
	\end{bmatrix} }_{\boldsymbol{F}(\theta,\phi)}   
\underbrace{\begin{bmatrix}
	\sin(\gamma)e^{j\eta} \\
	\cos(\gamma) \\
	\end{bmatrix}}_{\boldsymbol{g}(\gamma,\eta)} 
\end{aligned}    \label{11}
\end{equation}
where ${\boldsymbol{F}(\theta,\phi)}\in \mathbb{C}^{6\times 2}$ denotes the spatial angular location matrix, ${\boldsymbol{g}(\gamma,\eta)} \in \mathbb{C}^{2\times 1}$   {is representative of} the polarization states vector.
\par According to \cite{WongK}, the elevation angle $\theta$ and azimuth angle $\phi$ can be  extracted from the normalized Poynting vector of $\boldsymbol{q}(\theta,\phi,\gamma,\eta) $
\begin{equation}
\begin{aligned}
\begin{bmatrix}
{u}\\
{v}\\
{w}\\
\end{bmatrix}\triangleq \frac{\boldsymbol{q}_{[{1,2,3}]}}{\left \|\boldsymbol{q}_{[{1,2,3}]} \right \|} \times \frac{\boldsymbol{q}^{*}_{[{4,5,6}]}}{\left \|\boldsymbol{q}_{[{4,5,6}]} \right \|}=\begin{bmatrix}
\sin(\theta)\cos(\phi)\\
\sin(\theta)\sin(\phi)\\
\cos(\theta)\\
\end{bmatrix}
\end{aligned}    \label{12}
\end{equation}  
\par By using the orthogonality of the bistatic coprime EMVS-MIMO radar transmit and receive waveforms, the signal model after matched filtering at the receive front  can be {denoted} as
\begin{equation}
\begin{aligned}
\boldsymbol{y}(t)&=((\boldsymbol{A}_{t} \odot \boldsymbol{Q}_{t} )\odot (\boldsymbol{A}_{r} \odot \boldsymbol{Q}_{r} )) \boldsymbol{s}(t) + \boldsymbol{n}(t)\\
\end{aligned}    \label{13}
\end{equation}
where $\boldsymbol{A}_{t}=\begin{bmatrix}
\boldsymbol{a}_{t_{1}} &\boldsymbol{a}_{t_{2}}&\cdots&\boldsymbol{a}_{t_{K}}
\end{bmatrix}$ , $\boldsymbol{A}_{r}=\begin{bmatrix}
\boldsymbol{a}_{r_{1}} &\boldsymbol{a}_{r_{2}}&\cdots&\boldsymbol{a}_{r_{K}}
\end{bmatrix}$, $\boldsymbol{Q}_{t}=\begin{bmatrix}
\boldsymbol{q}_{t_{1}} &\boldsymbol{q}_{t_{2}}&\cdots&\boldsymbol{q}_{t_{K}}
\end{bmatrix}$ and $\boldsymbol{Q}_{r}=\begin{bmatrix}
\boldsymbol{q}_{r_{1}} &\boldsymbol{q}_{r_{2}}&\cdots&\boldsymbol{q}_{r_{K}}
\end{bmatrix}$ denote the transmit steering  matrix, the receive steering  matrix, the transmit  spatial response matrix  and the receive spatial response matrix, respectively. For the total $L$ snapshots, the sample data can be further {derived} as
\begin{equation}
\begin{aligned}
\boldsymbol{Y}
&=((\boldsymbol{A}_{t} \odot \boldsymbol{Q}_{t} )\odot (\boldsymbol{A}_{r} \odot \boldsymbol{Q}_{r} )) \boldsymbol{S} + \boldsymbol{N}
\end{aligned}   \label{14}
\end{equation}
\par For the bistatic coprime EMVS-MIMO radar with $M$-element transmit EMVS and $N$-element receive EMVS, the matrix dimension of $(\boldsymbol{C}_{t} \odot \boldsymbol{C}_{r})$ is $\mathbb{C}^{36MN\times K}$. In order to achieve the  underdetermined 2D-DOD and 2D-DOA,  the difference coarray of the transmit EMVS and receive EMVS {will be deployed}  to construct a new third-order tensor model. Moreover,  from (\ref{14}), it can be seen that the received data $\boldsymbol{Y}$ satisfies the space-time multi-dimensional tensor structure.
\par Here,  some assumptions {are made} in order to effectively estimate the  transmit 4-D parameter $(\boldsymbol{\theta}_{t},\boldsymbol{\phi}_{t},\boldsymbol{\gamma}_{t},\boldsymbol{\eta}_{t})$   and receive 4-D parameter $(\boldsymbol{\theta}_{r},\boldsymbol{\phi}_{r},\boldsymbol{\gamma}_{r},\boldsymbol{\eta}_{r})$,
\begin{itemize}
	\item The value of $(\boldsymbol{\theta}_{t},\boldsymbol{\phi}_{t},\boldsymbol{\gamma}_{t},\boldsymbol{\eta}_{t})$ and $(\boldsymbol{\theta}_{r},\boldsymbol{\phi}_{r},\boldsymbol{\gamma}_{r},\boldsymbol{\eta}_{r})$ should satisfy 
	$\boldsymbol{\theta}_{t},\boldsymbol{\theta}_{r}\in [0,\pi/2 )$, $\boldsymbol{\phi}_{t},\boldsymbol{\phi}_{r}\in [0,\pi/2 )$, $\boldsymbol{\gamma}_{t},\boldsymbol{\gamma}_{r}\in [0,\pi/2 )$, ,$\boldsymbol{\eta}_{t},\boldsymbol{\eta}_{r} \in [0,\pi/2 )$. 
	\item The signals are stationary, non-Gaussian random process and statistically independent of each other.
	\item The number of signals $K$ is konwn or corrected estimated by the existing algorithm.
	\item The $\boldsymbol{N}$ is zero-mean, complex Gaussian white noise.
	\item The noise is statistically independent of the signals. 
\end{itemize}

\section{The Proposed Difference Coarray Strategy}
\subsection{Construction of the Difference Coarray}
\par {For the estimation of}  more targets than elements, the difference coarray of transmit and receive EMVS are constructed.
From (\ref{9}) and (\ref{10}), it can be found that $\boldsymbol{a}_{t_{k}}$ and $\boldsymbol{q}_{t_{k}}$, $\boldsymbol{a}_{r_{k}}$ and $\boldsymbol{q}_{r_{k}}$ are copuling{, respectively}. {Therefore}, in order to obtain the difference coarrays, the de-coupling is invetiable.
\par {To} reserve the mutiple space-time structure in $\boldsymbol{Y}$, a 5-D tensor $\mathcal{Y}$ can be constructed as 
\begin{equation}
\begin{aligned}
\mathcal{Y}=\sum_{k=1}^{K}{\boldsymbol{{a}}_{tk}}\circ {\boldsymbol{{q}}_{tk}}\circ {\boldsymbol{{a}}_{rk}}\circ {\boldsymbol{{q}}_{rk}} \circ{\boldsymbol{{s}}_{k}} + \mathcal{N}_{l}
\end{aligned}   \label{15}
\end{equation}
\par Then, a 8-D tensor $\mathcal{R}$ can be obtained via the following correlation operation between $\mathcal{Y}$  and $\mathcal{Y}^{*}$ 
\begin{equation}
\begin{aligned}
\mathcal{R}&=\boldsymbol{E}[\mathcal{Y}\circ\mathcal{Y}^{*}]
=
\sum_{k=1}^{K}\sigma^{2}_{k}\underbrace{{\boldsymbol{{a}}_{tk}}}_1 \circ \underbrace{{\boldsymbol{{q}}_{tk}}}_2\circ \underbrace{{\boldsymbol{{a}}_{rk}}}_3\circ \underbrace{{\boldsymbol{{q}}_{rk}}}_4 \circ \underbrace{{\boldsymbol{{a}}^{*}_{tk}}}_5 \circ \underbrace{{\boldsymbol{{q}}^{*}_{tk}}}_6\circ \underbrace{{\boldsymbol{{a}}^{*}_{rk}}}_7\circ \underbrace{{\boldsymbol{{q}}^{*}_{rk}}}_8
+ \mathcal{R}_{N}
\end{aligned}   \label{16}
\end{equation}
where $\sigma^{2}_{k},k=1,2,\cdots,K$ denotes the power of the $k$-th source. And, according to $\textbf{Definition 1}$, the correspoding difference coarray of transmit EMVS and receive EMVS can be obtained by combining the factor vectors ${\boldsymbol{{a}}_{tk}}$ and ${\boldsymbol{{a}}^{*}_{tk}}$, ${\boldsymbol{{a}}_{rk}}$ and ${\boldsymbol{{a}}^{*}_{rk}}$, respectively.  
Thus, {the factor vectors can be rearranged as $[1,2,3,4,5,6,7,8]\mapsto [1,5,2,3,7,4,6,8]$ based on the tensor permutation rule, and }a new 8-D $\mathcal{R}_{1}$ can be expressed as
\begin{equation}
\begin{aligned}
\mathcal{R}_{1}&=\sum_{k=1}^{K}\sigma^{2}_{k}\underbrace{{\boldsymbol{{a}}_{tk}}}_1 \circ \underbrace{{\boldsymbol{{a}}^{*}_{tk}}}_5 \circ\underbrace{{\boldsymbol{{q}}_{tk}}}_2\circ\underbrace{{\boldsymbol{{a}}_{rk}}}_3\circ\underbrace{{\boldsymbol{{a}}^{*}_{rk}}}_7\circ
\underbrace{{\boldsymbol{{q}}_{rk}}}_4 \circ  \underbrace{{\boldsymbol{{q}}^{*}_{tk}}}_6\circ  \underbrace{{\boldsymbol{{q}}^{*}_{rk}}}_8
+ \mathcal{R}_{N1}\\
&=\sum_{k=1}^{K}\sigma^{2}_{k}\underbrace{{\boldsymbol{{a}}_{tk}}}_1 \circ \underbrace{{\boldsymbol{{a}}^{*}_{tk}}}_2 \circ\underbrace{{\boldsymbol{{q}}_{tk}}}_3\circ\underbrace{{\boldsymbol{{a}}_{rk}}}_4\circ\underbrace{{\boldsymbol{{a}}^{*}_{rk}}}_5\circ
\underbrace{{\boldsymbol{{q}}_{rk}}}_6 \circ  \underbrace{{\boldsymbol{{q}}^{*}_{tk}}}_7\circ  \underbrace{{\boldsymbol{{q}}^{*}_{rk}}}_8
+ \mathcal{R}_{N1}
\end{aligned}   \label{17}
\end{equation}
\par Then, the indeces $1$ and $2$,  $4$ and $5$,  $7$ and $8$ {need to be combined } by {the use of } the generalized tensorization of the canonical polyadic decomposition in $\textbf{Definition 7}$. As a result, a new 5-D tensor $\mathcal{R}_{2}=\mathcal{R}_{1{[1,2][4,5][7,8]}}$ can be {achieved} as 
\begin{equation}
\begin{aligned}
\mathcal{R}_{2}
&=\sum_{k=1}^{K}\sigma^{2}_{k}{\tilde{{\boldsymbol{{a}}}}}_{tk}  \circ{{\boldsymbol{{q}}_{tk}}}\circ{\tilde{{\boldsymbol{{a}}}}}_{rk} \circ
{{\boldsymbol{{q}}_{rk}}} \circ  {\tilde{{\boldsymbol{{q}}}}}_{k} 
+ \mathcal{R}_{N2}
\end{aligned}    \label{18}
\end{equation}
where ${\tilde{{\boldsymbol{{a}}}}}_{tk}={{\boldsymbol{{a}}_{tk}}} \otimes {{\boldsymbol{{a}}^{*}_{tk}}}$, ${\tilde{{\boldsymbol{{a}}}}}_{rk}={{\boldsymbol{{a}}_{rk}}}\otimes{{\boldsymbol{{a}}^{*}_{rk}}}$, ${\tilde{{\boldsymbol{{q}}}}}_{k} = {{\boldsymbol{{q}}^{*}_{tk}}}\otimes  {{\boldsymbol{{q}}^{*}_{rk}}}$.
\par According to $\textbf{Definition 1}$, for the $M$ transmit coprime EMVS with the number of the first subarray is $M1$ and the second subarray is $M2$, the $N$ transmit coprime EMVS with the number of the first subarray is $N1$ and the second subarray is $N2$,  the number of continue virtual array elements in the corresponding difference coarrays  can be deonted as 
\begin{equation}
\begin{aligned}
\widetilde{M}=2M_{1}M_{2}+2M_{1}+1
\end{aligned}   \label{19}
\end{equation}
\begin{equation}
\begin{aligned}
\widetilde{N}=2N_{1}N_{2}+2N_{1}+1
\end{aligned}   \label{20}
\end{equation}
\par {However,} the difference coarrays  of the coprime transmit array and the coprime receive array contain many duplicated elements. {To attain} the maximum central contiguous ULA in the difference coarrays, it is necessary to remove the repeated {elements, which can be realized by the application of tensor matrix product rules.} Two selection matrices $\boldsymbol{{J}}_{1}$ and $\boldsymbol{{J}}_{2}$ are designed {on the basis of}  $\textbf{Algorithm 1}$. {So}, a new 6-D tensor can be expressd as
\begin{equation}
\begin{aligned}
\mathcal{R}_{3} &= \mathcal{R}_{2}\times_{1}\boldsymbol{{J}}_{1}\times_{3}\boldsymbol{{J}}_{2}\\
&=\sum_{k=1}^{K}\sigma^{2}_{k}{\widetilde{{\boldsymbol{{a}}}}}_{tk}  \circ{{\boldsymbol{{q}}_{tk}}}\circ{\widetilde{{\boldsymbol{{a}}}}}_{rk} \circ
{{\boldsymbol{{q}}_{rk}}} \circ  {\tilde{{\boldsymbol{{q}}}}}_{k} 
+ \mathcal{R}_{N2}
\end{aligned}   \label{21}
\end{equation}
where 
\begin{equation}
\begin{aligned}
\boldsymbol{\widetilde{a}}_{tk}= \begin{bmatrix}
e^{-j \pi \frac{\widetilde{M}-1}{2} \sin(\theta_{t_{k}})} 
&\cdots&e^{j\pi \frac{\widetilde{M}-1}{2}  \sin(\theta_{t_{k}})}
\end{bmatrix}^{T}
\end{aligned}   \label{22}
\end{equation}
\begin{equation}
\begin{aligned}
\boldsymbol{\widetilde{a}}_{rk}= \begin{bmatrix}
e^{-j\pi \frac{\widetilde{N}-1}{2} \sin(\theta_{t_{r}})} 
&\cdots&e^{j\pi \frac{\widetilde{N}-1}{2}  \sin(\theta_{r_{k}})}
\end{bmatrix}^{T}
\end{aligned}   \label{23}
\end{equation}
\par It can be found that both the transmit {and}  the receive difference coarray can be {achievable} via the {above-mentioned} above process. Compared with (\ref{15}), it {is obvious} that the proposed process has significantly improved the degree of freedoms. 
\begin{algorithm}  
	\caption{pseudocode for coarray selection matrix }  
	\begin{algorithmic}[1]   
		\Require sparse array  $\mathbb{S}$, difference coarray
		$\mathbb{D}$, weight function $\mathbb{W}$
		\Ensure coarray selection matrix $\boldsymbol{J}$  
		\State construct rectangular grid $\boldsymbol{S}_{1}$ and $\boldsymbol{S}_{2} $ via $[\boldsymbol{S}_{1},\boldsymbol{S}_{2}]= ndgrid(\mathbb{S})$
		\State $\boldsymbol{S}_{3}= \boldsymbol{S}_{1}-\boldsymbol{S}_{2}$
		\State   $\boldsymbol{S}_{4}=vec(\boldsymbol{S}_{3})$
		\State $\boldsymbol{J} = zeros(|\mathbb{D}|,|\boldsymbol{S}_{3}|) $ 
		\State $\boldsymbol{for} \quad i=1:|\mathbb{D}|  $   
		\State  \quad \quad \quad $[row,col] = find(\boldsymbol{S}_{4}==\mathbb{D}(i))$
		\State \quad \quad \quad $\boldsymbol{J}(i,row) = 1/\mathbb{W}(i)$
		\State \Return{$\boldsymbol{J}$}  
	\end{algorithmic}  
\end{algorithm} 
\subsection{Formulation of the new third tensor}
It can be seen that $\boldsymbol{\widetilde{a}}_{tk}$ and $\boldsymbol{\widetilde{a}}_{rk}$ in (\ref{22}) and (\ref{23}) satisfy the {centro}-symmetric {geometry, so},  the DFT beamspace matrices are used to 
transform the complex value of $\boldsymbol{\widetilde{a}}_{tk}$ and $\boldsymbol{\widetilde{a}}_{rk}$ into the corresponding real-value ones. {This operation can improve}  the estimation performance of the angle parameters and {help}  to reconstruct the spatial response matrices $\boldsymbol{Q}_{t}$ and $\boldsymbol{Q}_{r}$ {easily}. Firstly, two DFT beamspace matrices are designed as
\begin{equation}
\begin{aligned}
\boldsymbol{W}_{t}&=
\begin{bmatrix}
\boldsymbol{w}^{H}_{t1}\\
\boldsymbol{w}^{H}_{t2}\\
\vdots\\
\boldsymbol{w}^{H}_{t\widetilde{M}}\\
\end{bmatrix}\\
\end{aligned}  \label{24} 
\end{equation}
\begin{equation}
\begin{aligned}
\boldsymbol{W}_{r}&=
\begin{bmatrix}
\boldsymbol{w}^{H}_{r1}\\
\boldsymbol{w}^{H}_{r2}\\
\vdots\\
\boldsymbol{w}^{H}_{r\widetilde{N}}\\
\end{bmatrix}
\end{aligned}   \label{25} 
\end{equation}
where
\begin{equation}
\begin{aligned}
\boldsymbol{w}_{tm}&=
\begin{bmatrix}
e^{-j \frac{\pi(\widetilde{M}-1)\boldsymbol{u}_{t}(m)}{\widetilde{M}} } 
&\cdots&e^{j \frac{\pi(\widetilde{M}-1)\boldsymbol{u}_{t}(m)}{\widetilde{M}} } 
\end{bmatrix}^{T}\\
&\quad \quad \quad\quad \quad m=1,\cdots,\widetilde{M}
\end{aligned}   \label{26} 
\end{equation}
\begin{equation}
\begin{aligned}
\boldsymbol{w}_{rn}&=\begin{bmatrix}
e^{-j\frac{\pi(\widetilde{N}-1)\boldsymbol{u}_{r}(n)}{\widetilde{N}} } 
&\cdots&e^{j\frac{\pi(\tilde{N}-1)\boldsymbol{u}_{r}(n)}{\widetilde{N}} } 
\end{bmatrix}^{T}\\
&\quad \quad \quad\quad \quad n=1,\cdots,\widetilde{N}
\end{aligned}   \label{27} 
\end{equation}
where $\boldsymbol{u}_{t}=[0,1,\cdots,\widetilde{M}-1]$, $\boldsymbol{u}_{r}=[0,1,\cdots,\widetilde{N}-1]$. Then,
a 5-D tensor with real-value steering matrices can be
obtained by exploiting the  tensor-matrix
product rules 
\begin{equation}
\begin{aligned}
\mathcal{R}_{4} &= \mathcal{R}_{3}\times_{1}\boldsymbol{W}_{t}\times_{3}\boldsymbol{W}_{r}\\
&=\sum_{k=1}^{K}\sigma^{2}_{k}{\widehat{{\boldsymbol{{a}}}}}_{tk}  \circ{{\boldsymbol{{q}}_{tk}}}\circ{\widehat{{\boldsymbol{{a}}}}}_{rk} \circ
{{\boldsymbol{{q}}_{rk}}} \circ  {\tilde{{\boldsymbol{{q}}}}}_{k} 
+ \mathcal{R}_{N4}
\end{aligned}   \label{28}
\end{equation}
where  $\boldsymbol{\widehat{A}}_{t}=\boldsymbol{W}_{t}\boldsymbol{\widetilde{A}}_{t}=[\boldsymbol{\widehat{a}}_{t1},\boldsymbol{\widehat{a}}_{t2},\cdots,\boldsymbol{\widehat{a}}_{tK}] $ and $\boldsymbol{\widehat{A}}_{r}=\boldsymbol{W}_{r}\boldsymbol{\widetilde{A}}_{r}=[\boldsymbol{\widehat{a}}_{r1},\boldsymbol{\widehat{a}}_{r2},\cdots,\boldsymbol{\widehat{a}}_{rK}] $ denote the real-value steering vector matrices of the transmit array and the receive array, respectively, and their detailed {forms} are as (\ref{29}) and (\ref{30}).
\begin{figure*}[htbp]
	\begin{equation}
	\label{29}
	\boldsymbol{\widehat{A}}_{t}
	=
	\begin{bmatrix}
	\frac{sin(\pi\frac{ \widetilde{M} (sin\theta_{t_{1}})}{2})}{sin(\frac{(sin\theta_{t_{1}})}{2})}
	&\frac{sin(\pi\frac{ \widetilde{M} (sin\theta_{t_{2}})}{2})}{sin(\frac{(sin\theta_{t_{2}})}{2})} &\cdots&\frac{sin(\pi\frac{ \widetilde{M} (sin\theta_{t_{K}})}{2})}{sin(\pi\frac{(sin\theta_{t_{K}})}{2})} \\
	\frac{sin(\pi\frac{ \widetilde{M} (sin\theta_{t_{1}}-\frac{2\pi}{\widetilde{M}})}{2})}{sin(\pi\frac{(sin\theta_{t_{1}}-\frac{2\pi}{\widetilde{M}})}{2})} & \frac{sin(\pi\frac{ \widetilde{M} (sin\theta_{t_{2}}-\frac{2\pi}{\widetilde{M}})}{2})}{sin(\pi\frac{(sin\theta_{t_{2}}-\frac{2\pi}{\widetilde{M}})}{2})}
	&\cdots& \frac{sin(\pi\frac{ \widetilde{M} (sin\theta_{t_{K}}-\frac{2\pi}{\widetilde{M}})}{2})}{sin(\pi\frac{(sin\theta_{t_{K}}-\frac{2\pi}{\widetilde{M}})}{2})}\\
	\vdots&\vdots&\ddots&\vdots\\		
	\frac{sin(\pi\frac{ \widetilde{M} (sin\theta_{t_{1}}-\frac{2(\widetilde{M}-1)\pi}{\widetilde{M}})}{2})}{sin(\pi\frac{(sin\theta_{t_{1}}-\frac{2(\widetilde{M}-1)\pi}{\widetilde{M}})}{2})}&\frac{sin(\pi\frac{ \widetilde{M} (sin\theta_{t_{2}}-\frac{2(\widetilde{M}-1)\pi}{\widetilde{M}})}{2})}{sin(\pi\frac{(sin\theta_{t_{2}}-\frac{2(\widetilde{M}-1)\pi}{\widetilde{M}})}{2})}&\cdots&\frac{sin(\frac{ \widetilde{M} (sin\theta_{t_{K}}-\frac{2(\widetilde{M}-1)\pi}{\widetilde{M}})}{2})}{sin(\pi\frac{(sin\theta_{t_{K}}-\frac{2(\widetilde{M}-1)\pi}{\widetilde{M}})}{2})}
	\end{bmatrix} 
	\end{equation} 
\end{figure*}
\begin{figure*}[htbp]
	\begin{equation}
	\label{30}
	\boldsymbol{\widehat{A}}_{r}
	=
	\begin{bmatrix}
	\frac{sin(\pi\frac{ \widetilde{N} (sin\theta_{r_{1}})}{2})}{sin(\pi\frac{(sin\theta_{r_{1}})}{2})}
	&\frac{sin(\pi\frac{ \widetilde{N} (sin\theta_{r_{2}})}{2})}{sin(\pi\frac{(sin\theta_{r_{2}})}{2})} &\cdots&\frac{sin(\pi\frac{ \widetilde{N} (sin\theta_{r_{K}})}{2})}{sin(\pi\frac{(sin\theta_{r_{K}})}{2})} \\
	\frac{sin(\pi\frac{ \widetilde{N} (sin\theta_{r_{1}}-\frac{2\pi}{\widetilde{N}})}{2})}{sin(\pi\frac{(sin\theta_{r_{1}}-\frac{2\pi}{\widetilde{N}})}{2})} & \frac{sin(\pi\frac{ \widetilde{N} (sin\theta_{r_{2}}-\frac{2\pi}{\widetilde{N}})}{2})}{sin(\pi\frac{(sin\theta_{r_{2}}-\frac{2\pi}{\widetilde{N}})}{2})}
	&\cdots& \frac{sin(\pi\frac{ \widetilde{N} (sin\theta_{r_{K}}-\frac{2\pi}{\widetilde{N}})}{2})}{sin(\pi\frac{(sin\theta_{r_{K}}-\frac{2\pi}{\widetilde{N}})}{2})}\\
	\vdots&\vdots&\ddots&\vdots\\		
	\frac{sin(\pi\frac{ \widetilde{N} (sin\theta_{r_{1}}-\frac{2(\widetilde{N}-1)\pi}{\widetilde{N}})}{2})}{sin(\pi\frac{(sin\theta_{r_{1}}-\frac{2(\widetilde{N}-1)\pi}{\widetilde{N}})}{2})}&\frac{sin(\pi\frac{ \widetilde{N} (sin\theta_{r_{2}}-\frac{2(\widetilde{N}-1)\pi}{\widetilde{N}})}{2})}{sin(\pi\frac{(sin\theta_{r_{2}}-\frac{2(\widetilde{N}-1)\pi}{\widetilde{N}})}{2})}&\cdots&\frac{sin(\pi\frac{ \widetilde{N} (sin\theta_{r_{K}}-\frac{2(\widetilde{N}-1)\pi}{\widetilde{N}})}{2})}{sin(\pi\frac{(sin\theta_{r_{K}}-\frac{2(\widetilde{N}-1)\pi}{\widetilde{N}})}{2})}
	\end{bmatrix}  
	\end{equation} 
\end{figure*}
\par {Afterwards}, a new three way tensor $\mathcal{R}_{5}$ can be {attained} via combining the {indices}  1 and 2, 3 and 4 
\begin{equation}
\begin{aligned}
\mathcal{R}_{5} = \mathcal{R}_{4[1,2][3,4][5]}= \sum_{k=1}^{K}{\boldsymbol{\widehat{c}}_{tk}}\circ {\boldsymbol{\widehat{c}}_{rk}} \circ{\boldsymbol{\widehat{q}}_{k}}+ \mathcal{R}_{N5}
\end{aligned}     \label{31} 
\end{equation}
\begin{figure}[htbp]
	\centerline{\includegraphics[width=26.5pc]{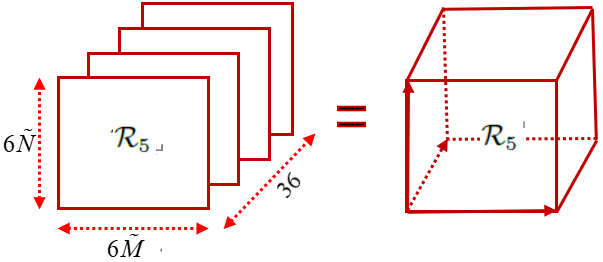}}
	\caption{The three-way tensor $\mathcal{R}_{5}$. }
	\label{tensor2}
\end{figure}
\par As shown in  Fig. \ref{tensor2}, the dimension of the {newly} constructed tensor $\mathcal{R}_{5}$ is $\mathbb{C}^{6\widetilde{M}\times 6\widetilde{N}\times 36}${, whose third way is fixed at 36.} 
{On its basis,} the PARAFAC algorithm can be {utilized}  to estimate the   factor matrices  ${\boldsymbol{\widehat{C}}_{t}}$, ${\boldsymbol{\widehat{C}}_{r}}$, and ${\boldsymbol{\widehat{Q}}}$, respectively. According to $\textbf{Definition 6}$, the different  {slices} of  the  third-order tensor $\mathcal{R}_{5}$ can be further {obtained} as 
\begin{equation}
\begin{aligned}
\mathcal{R}^{T}_{5[{i,:,:}]} &= {\boldsymbol{\widehat{C}}_{r}} \mathcal{D}_{i}({{\boldsymbol{\widehat{Q}}}} )  {\boldsymbol{\widehat{C}}_{t}}^{T},\quad i =1,2,\cdots,36 \\
\mathcal{R}^{T}_{5[{:,j,:}]} &={\boldsymbol{\widehat{Q}}}\mathcal{D}_{j}({{\boldsymbol{\widehat{C}}_{t}}} )  {\boldsymbol{\widehat{C}}_{r}}^{T}  ,\quad j=1,2,\cdots,6\widetilde{M}\\
\mathcal{R}^{T}_{5[{:,:,k}]} &= {\boldsymbol{\widehat{C}}_{t}} \mathcal{D}_{k}({\boldsymbol{\widehat{C}}_{r}})  {\boldsymbol{\widehat{Q}}}^{T},\quad k=1,2,\cdots,6\widetilde{N}\\
\end{aligned}  \label{32} 
\end{equation}
where $\mathcal{D}_{i}$, $\mathcal{D}_{j}$, $\mathcal{D}_{k}$ represent the diagonal matrix operation, and the elements on the diagonal matrices are the {$k$th}-rows of the factor matrices ${\boldsymbol{\widehat{C}}_{t}}$, ${{\boldsymbol{\widehat{C}}_{r}}}$, and ${\boldsymbol{\widehat{Q}}}$, respectively.
\par In order to realize the estimation of transmit steering  matrix ${\boldsymbol{\widehat{C}}_{t}}$, the receive steering  matrix ${{\boldsymbol{\widehat{C}}_{r}}}$ and the signal matrix ${\boldsymbol{\widehat{Q}}}$, the trilinear alternating least squares algorithm is {deployed as follows} 
\begin{equation}
\begin{aligned}
\min_{{\boldsymbol{\widehat{C}}^{T}_{t}}}&=\Vert \lbrack \mathcal{R}^{T}_{5} \rbrack_{(1)} -  \left ( {{\boldsymbol{\widehat{C}}_{r}}}  \odot  {\boldsymbol{\widehat{Q}}} \right ){\boldsymbol{\widehat{C}}^{T}_{t}}   \Vert^{2}_{F}\\
\min_{{\boldsymbol{\widehat{C}}^{T}_{r}}}&=\Vert \lbrack  \mathcal{R}^{T}_{5}\rbrack_{(2)} - \left ( {\boldsymbol{\widehat{Q}}}\odot  {{\boldsymbol{\widehat{C}}_{t}}}  \right )  {\boldsymbol{\widehat{C}}^{T}_{r}}  \Vert^{2}_{F}\\
\min_{{\boldsymbol{\widehat{Q}}^{T}}}&=\Vert \lbrack  \mathcal{R}^{T}_{5}\rbrack_{(3)} - \left ( {\boldsymbol{\widehat{C}}_{t}}\odot  {\boldsymbol{\widehat{C}}_{r}} \right )  {\boldsymbol{\widehat{Q}}}^{T}    \Vert^{2}_{F}\\
\end{aligned}   \label{33} 
\end{equation}
\par Then, let ${{\boldsymbol{\breve{Q}}}}$, ${\boldsymbol{\breve{C}}_{t}}$ and ${\boldsymbol{\breve{C}}_{r}}$ denote the estimated factor matrices, respectively
\begin{equation}
\begin{aligned}
&{\boldsymbol{\breve{C}}^{T}_{t}}= (\boldsymbol{\widehat{C}}_{r}\odot \boldsymbol{\widehat{Q}})^{\dagger}\lbrack  \mathcal{R}^{T}_{5}\rbrack_{(1)} \\
&{{\boldsymbol{\breve{C}}^{T}_{r}}}= ({\boldsymbol{\widehat{Q}}}\odot \boldsymbol{\widehat{C}}_{t})^{\dagger}\lbrack  \mathcal{R}^{T}_{5}\rbrack_{(2)} \\
&{{\boldsymbol{\breve{Q}}}}^{T} = (\boldsymbol{\widehat{C}}_{t}\odot \boldsymbol{\widehat{C}}_{r})^{\dagger}\lbrack  \mathcal{R}^{T}_{5}\rbrack_{(3)} \\
\end{aligned}   \label{34} 
\end{equation}
\par  For the factor matrices ${{\boldsymbol{\breve{Q}}}}$, ${\boldsymbol{\breve{C}}_{t}}$ and ${\boldsymbol{\breve{C}}_{r}}$ obtained by multiple iterations of PARAFAC-TALS, the transmit steering  matrices ${\boldsymbol{\breve{C}}_{t}}$ and the receive steering  matrices ${\boldsymbol{\breve{C}}_{r}}$ are in one-to-one correspondence. This means that the 2D-DOD and 2D-DOA in ${\boldsymbol{\breve{C}}_{t}}$ and ${\boldsymbol{\breve{C}}_{r}}$ are automatically paired. Therefore, the proposed algorithm {need not the construction of} an additional pairing optimization function compared to the algorithms in \cite{Chintagunta}-\cite{MaoC}. 
\subsection{Estimaiton of the angle and polarization parameters}
\par 
For the estimated ${\boldsymbol{\breve{C}}_{t}}$ and ${\boldsymbol{\breve{C}}_{r}}$, the corresponding transmit elevation angle  and receive elevation angle can be obtained. Firstly, structural characteristics of the steering vectors  $\boldsymbol{\widehat{a}}_{tk}$ in $\boldsymbol{\widehat{A}}_{t}$ and  $\boldsymbol{\widehat{a}}_{rk}$ in $\boldsymbol{\widehat{A}}_{r}$ { are analyzed}. The $m$-th and $(m+1)$-th elements in $\boldsymbol{\widehat{a}}_{tk}$  and $n$-th and $(n+1)$-th  elements in $\boldsymbol{\widehat{a}}_{rk}$ can be expressed as
\begin{equation}
\begin{aligned}
\begin{cases}
\lbrack\boldsymbol{\widehat{a}}_{tk} \rbrack_{m}= \frac{sin(\pi\frac{ \widetilde{M} (sin\theta_{t_{k}}-\frac{2m\pi}{\widetilde{M}})}{2})}{sin(\pi\frac{(sin\theta_{t_{k}}-\frac{2m\pi}{\widetilde{M}})}{2})}   \\
\lbrack\boldsymbol{\widehat{a}}_{tk} \rbrack_{m+1}= \frac{sin(\pi\frac{ \widetilde{M} (sin\theta_{t_{k}}-\frac{2(m+1)\pi}{\widetilde{M}})}{2})}{sin(\pi\frac{(sin\theta_{t_{k}}-\frac{2(m+1)\pi}{\widetilde{M}})}{2})}\\
\end{cases}\\ 
\end{aligned}   
\end{equation}
\begin{equation}
\begin{aligned}
\begin{cases}
\lbrack\boldsymbol{\widehat{a}}_{rk} \rbrack_{n}= \frac{sin(\pi\frac{ \widetilde{N} (sin\theta_{r_{k}}-\frac{2n\pi}{\widetilde{N}})}{2})}{sin(\pi\frac{(sin\theta_{r_{k}}-\frac{2m\pi}{\widetilde{N}})}{2})}   \\
\lbrack\boldsymbol{\widehat{a}}_{rk} \rbrack_{n+1}= \frac{sin(\pi\frac{ \widetilde{N} (sin\theta_{r_{k}}-\frac{2(n+1)\pi}{\widetilde{N}})}{2})}{sin(\pi\frac{(sin\theta_{r_{k}}-\frac{2(n+1)\pi}{\widetilde{N}})}{2})}\\
\end{cases}\\ 
\end{aligned}   
\end{equation}
\par It can be seen that the adjacent elements in  $\boldsymbol{\widehat{a}}_{tk} $ and $\boldsymbol{\widehat{a}}_{rk} $ satisfy the following  {relationships}
\begin{equation}
\begin{aligned}
&tan(\frac{\pi sin(\theta_{t_{k}})}{2})\lbrace cos(\frac{m\pi}{{\widetilde{M}}})\lbrack\boldsymbol{\widehat{a}}_{tk} \rbrack_{m} +cos (\frac{(m+1)\pi}{{\widetilde{M}}})\lbrack\boldsymbol{\widehat{a}}_{tk} \rbrack_{m+1} \rbrace\\
&\quad\quad\quad\quad =sin(\frac{m\pi}{{\widetilde{M}}})\lbrack\boldsymbol{\widehat{a}}_{tk} \rbrack_{m} +sin (\frac{(m+1)\pi}{{\widetilde{M}}})\lbrack\boldsymbol{\widehat{a}}_{tk} \rbrack_{m+1} 
\end{aligned}   \label{37}
\end{equation}
\begin{equation}
\begin{aligned}
&tan(\frac{\pi sin(\theta_{r_{k}})}{2})\lbrace cos(\frac{n\pi}{{\widetilde{N}}})\lbrack\boldsymbol{\widehat{a}}_{rk} \rbrack_{n} +cos (\frac{(n+1)\pi}{{\widetilde{N}}})\boldsymbol\lbrack\boldsymbol{\widehat{a}}_{rk} \rbrack_{n+1} \rbrace\\
&\quad\quad\quad\quad =sin(\frac{n\pi}{{\widetilde{N}}})\lbrack\boldsymbol{\widehat{a}}_{rk} \rbrack_{n} +sin (\frac{(n+1)\pi}{{\widetilde{N}}})\boldsymbol\lbrack\boldsymbol{\widehat{a}}_{rk} \rbrack_{n+1} 
\end{aligned}   \label{38}
\end{equation}
\par {On the basis of} (\ref{37}) and (\ref{38}),  the following rotation-invariant relationship for $\boldsymbol{\widehat{A}}_{t}$ and $\boldsymbol{\widehat{A}}_{r}$ can be constructed
\begin{equation}
\begin{aligned}
\boldsymbol{\Gamma}_{t1}\boldsymbol{\widehat{A}}_{t}\boldsymbol{\Phi}(\theta_{t})=\boldsymbol{\Gamma}_{t2}\boldsymbol{\widehat{A}}_{t}
\end{aligned}   
\end{equation}
\begin{equation}
\begin{aligned}
\boldsymbol{\Gamma}_{r1}\boldsymbol{\widehat{A}}_{r}\boldsymbol{\Phi}(\theta_{r})=\boldsymbol{\Gamma}_{r2}\boldsymbol{\widehat{A}}_{r}
\end{aligned}   
\end{equation}
where
\begin{equation}
\begin{aligned}
\boldsymbol{\Phi}(\theta_{t})=\begin{bmatrix}
tan(\frac{\pi sin(\theta_{t_{1}})}{2})&&\\
&\ddots& \\
&&tan(\frac{\pi sin(\theta_{t_{K}})}{2})\end{bmatrix}
\end{aligned}   
\end{equation}
\begin{equation}
\begin{aligned}
\boldsymbol{\Phi}(\theta_{r})=\begin{bmatrix}
tan(\frac{\pi sin(\theta_{r_{1}})}{2})&&\\
&\ddots& \\
&&tan(\frac{\pi sin(\theta_{r_{K}})}{2})\end{bmatrix}
\end{aligned}   
\end{equation}
\par The detailed {forms} of the selection matrices $\boldsymbol{\Gamma}_{t1}$,  $\boldsymbol{\Gamma}_{t2}$, $\boldsymbol{\Gamma}_{r1}$, and $\boldsymbol{\Gamma}_{r2}$ are
\begin{equation}
\begin{aligned}
\boldsymbol{\Gamma}_{t1}=
\begin{bmatrix}
{c}_{t1}&{c}_{t2}&0&\cdots&0\\
0&{c}_{t2}&{c}_{t3}&\cdots&0\\
0&0&{c}_{t3}&\cdots&0\\
\vdots&\vdots&\vdots&\ddots&\vdots\\
&0&0&\cdots&{c}_{t(\widetilde{M}-1)}
\end{bmatrix}
\end{aligned}  \label{43} 
\end{equation}
\begin{equation}
\begin{aligned}
\boldsymbol{\Gamma}_{t2}=
\begin{bmatrix}
{s}_{t1}&{s}_{t2}&0&\cdots&0\\
0&{s}_{t2}&{s}_{t3}&\cdots&0\\
0&0&{s}_{t3}&\cdots&0\\
\vdots&\vdots&\vdots&\ddots&\vdots\\
&0&0&\cdots&{s}_{t(\widetilde{M}-1)}
\end{bmatrix}
\end{aligned}    \label{44} 
\end{equation}
\begin{equation}
\begin{aligned}
\boldsymbol{\Gamma}_{r1}=
\begin{bmatrix}
{c}_{r1}&{c}_{r2}&0&\cdots&0\\
0&{c}_{r2}&{c}_{r3}&\cdots&0\\
0&0&{c}_{r3}&\cdots&0\\
\vdots&\vdots&\vdots&\ddots&\vdots\\
&0&0&\cdots&{c}_{r(\widetilde{N}-1)}
\end{bmatrix}
\end{aligned}    \label{45} 
\end{equation}
\begin{equation}
\begin{aligned}
\boldsymbol{\Gamma}_{r2}=
\begin{bmatrix}
{s}_{r1}&{s}_{r2}&0&\cdots&0\\
0&{s}_{r2}&{s}_{r3}&\cdots&0\\
0&0&{s}_{r3}&\cdots&0\\
\vdots&\vdots&\vdots&\ddots&\vdots\\
&0&0&\cdots&{s}_{r(\widetilde{N}-1)}
\end{bmatrix}
\end{aligned}    \label{46} 
\end{equation}
where ${c}_{tm}=cos(\frac{m\pi}{\widetilde{M}}),m=0,1,\cdots,\widetilde{M}$, ${c}_{rn}=cos(\frac{n\pi}{\widetilde{N}}),n=0,1,\cdots,\widetilde{N}$, ${s}_{tm}=sin(\frac{m\pi}{\widetilde{M}}),m=0,1,\cdots,\widetilde{M}$, ${s}_{rm}=sin(\frac{n\pi}{\widetilde{N}}),n=0,1,\cdots,\widetilde{N}$. Therefore, in order to estimate the  transmit elevation angle $\theta_{t_{k}},k=1,2,\cdots,K$ and the receive elevation angle $\theta_{r_{k}},k=1,2,\cdots,K$ in the bistatic EMVS-MIMO radar system, the following rotation-invariant relationship is further constructed 
\begin{equation}
\begin{aligned}
(\boldsymbol{\Gamma}_{t2}\otimes \boldsymbol{I}_{6}){\boldsymbol{\breve{C}}_{t}}=(\boldsymbol{\Gamma}_{t1}\otimes \boldsymbol{I}_{6}){\boldsymbol{\breve{C}}_{t}}
\boldsymbol{\Phi}(\theta_{t})
\end{aligned}    \label{47} 
\end{equation}
\begin{equation}
\begin{aligned}
(\boldsymbol{\Gamma}_{r2}\otimes \boldsymbol{I}_{6}){\boldsymbol{\breve{C}}_{r}}=(\boldsymbol{\Gamma}_{r1}\otimes \boldsymbol{I}_{6}){\boldsymbol{\breve{C}}_{r}}
\boldsymbol{\Phi}(\theta_{r})
\end{aligned}    \label{48} 
\end{equation}
\par Then, the estimation of $\boldsymbol{\Phi}(\theta_{t})$ and $\boldsymbol{\Phi}(\theta_{r})$ can be achieved by using the least squares 
\begin{equation}
\begin{aligned}
\boldsymbol{\breve{\Phi}}(\theta_{t})=((\boldsymbol{\Gamma}_{t1}\otimes\boldsymbol{I}_{6}){\boldsymbol{\breve{C}}_{t}})^{\dagger}(\boldsymbol{\Gamma}_{t2}\otimes \boldsymbol{I}_{6}){\boldsymbol{\breve{C}}_{t}}
\end{aligned}     \label{49} 
\end{equation}
\begin{equation}
\begin{aligned}
\boldsymbol{\breve{\Phi}}(\theta_{r})=((\boldsymbol{\Gamma}_{r1}\otimes\boldsymbol{I}_{6}){\boldsymbol{\breve{C}}_{r}})^{\dagger}(\boldsymbol{\Gamma}_{r2}\otimes \boldsymbol{I}_{6}){\boldsymbol{\breve{C}}_{r}}
\end{aligned}    \label{50} 
\end{equation}
\par {Furthermore}, by performing the singular value decomposition on $\boldsymbol{\breve{\Phi}}(\theta_{t})$ and $\boldsymbol{\breve{\Phi}}(\theta_{r})$, the corresponding eigenvalues $\lambda_{t1},\lambda_{t2},\cdots,\lambda_{tK}$ and $\lambda_{r1},\lambda_{r2},\cdots,\lambda_{rK}$ can be {attained}. Thus, the  transmit elevation angle $\theta_{tk},k=1,2,\cdots,K$ and receive elevation angle $\theta_{rk},k=1,2,\cdots,K$ can be estimated as
\begin{equation}
\begin{aligned}
\breve{\theta}_{tk}=arcsin(\frac{2arctan(\lambda_{tk})}{\pi}),\quad k=1,2,\cdots,K
\end{aligned}   
\end{equation}
\begin{equation}
\begin{aligned}
\breve{\theta}_{rk}=arcsin(\frac{2arctan(\lambda_{rk})}{\pi}),\quad k=1,2,\cdots,K
\end{aligned}   
\end{equation}
\par For the estimated $\breve{\theta}_{tk}, k=1,2,\cdots,K$ and $\breve{\theta}_{rk},k=1,2,\cdots,K$, the corresponding transmit/receive azimuth angle, transmit/receive polarization angle and transmit/receive polarization phase difference can be {gained} by the following process.  Since the estimation process for $(\phi_{t_{k}},\gamma_{t_{k}},\eta_{t_{k}}),k=1,2,\cdots,K$ and $(\phi_{r_{k}},\gamma_{r_{k}},\eta_{r_{k}}),k=1,2,\cdots,K$ are similar, this paper only {provides}  the comprehensive derivation process for $(\phi_{t_{k}},\gamma_{t_{k}},\eta_{t_{k}}),k=1,2,\cdots,K$.
\par By exploiting the property of Khatri-Rao product, the transmit spatial response $\boldsymbol{Q}_{t}$ can be reconstructed as 
\begin{equation}
\begin{aligned}
\begin{bmatrix}
\boldsymbol{q}_{t1},\cdots,\boldsymbol{q}_{t1}
\end{bmatrix}=\frac{1}{{\widetilde{M}}}\sum_{m=1}^{{\widetilde{M}}}({\boldsymbol{\breve{C}}_{t}}(6m-5:6m,:))
\end{aligned}   \label{53}
\end{equation}
where $\boldsymbol{\breve{C}}_{t}(6m-5:6m,:)$ denote $(6m-5)$-th row to $6m$-th row of $\boldsymbol{\breve{C}}_{t}$. Due to the real-value transform in (\ref{28}), there is no need to multiply an additional diagonal matrix when estimating the $\boldsymbol{Q}_{t}${, which differs from the reconstruction of $\boldsymbol{Q}_{t}$ in \cite{Chintagunta}-\cite{XianpengWang}.}.
\par For the reconstructed transmit spatial response $\boldsymbol{Q}_{t}$ ,  the normalized Poynting vector corresponding to the transmit EMVS array  {is}
\begin{equation}
\begin{aligned}
\begin{bmatrix}
\tilde{u}_{t_{k}}\\
\tilde{v}_{t_{k}}\\
\tilde{w}_{t_{k}}\\
\end{bmatrix}\triangleq \frac{\boldsymbol{e}_{t_{k}}}{\left \|\boldsymbol{e}_{t_{k}}  \right \|} \times \frac{\boldsymbol{h}^{*}_{t_{k}}}{\left \|\boldsymbol{h}_{t_{k}}  \right \|}=\begin{bmatrix}
\sin(\theta_{t_{k}})\cos(\phi_{t_{k}})\\
\sin(\theta_{t_{k}})\sin(\phi_{t_{k}})\\
\cos(\theta_{t_{k}})\\
\end{bmatrix}
\end{aligned}   \label{54}
\end{equation} 
\par Therefore, the estimated transmit azimuth angle $\boldsymbol{\phi}_{t_{k}},k=1,2,\cdots,K$ can be expressed as 
\begin{equation}
\begin{aligned}
\breve{{\phi}}_{t_{k}}=arctan(\frac{ \tilde{v}_{t_{k}}}{ \tilde{u}_{t_{k}}}),\quad k=1,2,\cdots,K
\end{aligned}  \label{55} 
\end{equation}
\par After obtaining $({\theta}_{t_{k}},{\phi}_{t_{k}}),k=1,2,\cdots,K$, the corresponding transmit  polarization state vector ${\boldsymbol{g}_{t_{k}}(\gamma_{t_{k}},\eta_{t_{k}})} $ can be {achieved}
\begin{equation}
\begin{aligned}
{\boldsymbol{g}_{t_{k}}(\gamma_{t_{k}},\eta_{t_{k}})}&=
\begin{bmatrix}
{\boldsymbol{g}_{1t_{k}}}\\
{\boldsymbol{g}_{2t_{k}}}\\
\end{bmatrix}=[{\boldsymbol{F}(\breve{\theta}_{t_{k}},\breve{\phi}_{t_{k}})} ]^{\dagger}\boldsymbol{Q}_{t},\\
&	{ k=1,2,\cdots,K
}\end{aligned}   \label{56}
\end{equation}
\par {Consequently}, the polarization parameters $(\gamma_{t_{k}},\eta_{t_{k}}), k=1,2,\cdots,K$ of the transmit EMVS can be {obtained} as 
\begin{equation}
\begin{array}{lcl}
\breve{\gamma}_{t_{k}}=arctan[\frac{{\boldsymbol{g}_{1t_{k}}}}{{\boldsymbol{g}_{2t_{k}}}}]  \\
\breve{\eta}_{t_{k}}=\angle{\boldsymbol{g}_{1t_{k}}}	 
\end{array},k=1,2,\cdots,K  \label{57}
\end{equation}
\par Finally,  the automatically paired transmit 4-D parameters $(\theta_{t},\phi_{t},\gamma_{t},\eta_{t}),k=1,2,\cdots,K$ can be {gained} by using {the mentioned above process, and the receive } 4-D parameters $(\theta_{r},\phi_{r},\gamma_{r},\eta_{r}),k=1,2,\cdots,K$ can be estimated in the same way. $\textbf{Algorithm 2}$ {offers} the framework of the proposed algorithm.
\begin{algorithm}  
	\caption{Framework of the proposed algorithm }  
	\begin{algorithmic}[1]   
		\Require array received data matrix $\boldsymbol{Y}$
		\Ensure the transmit 4-D parameters $(\boldsymbol{\theta}_{t},\boldsymbol{\phi}_{t},\boldsymbol{\gamma}_{t},\boldsymbol{\eta}_{t})$ and the receive 4-D parameters $(\boldsymbol{\theta}_{r},\boldsymbol{\phi}_{r},\boldsymbol{\gamma}_{r},\boldsymbol{\eta}_{r})$  
		\State construct the 5-D tensor $\mathcal{Y}$ according to (\ref{15}) 
		\State compute the  covariance tensor $\mathcal{R}$ according to (\ref{16}) 
		\State   calculate the new 5-D tensor $\mathcal{R}_{2}$ by using the tensor permutation and the generalized tensorization of the canonical polyadic decomposition according to (\ref{17}) and (\ref{18})
		\State remove the repeated elements in the difference coarrays in $\mathcal{R}_{2}$ according to (\ref{21})
		\State construct the three-way tensor  $\mathcal{R}_{5}$ according to (\ref{24})-(\ref{31})
		\State  perform the PARAFAC algorithm on  $\mathcal{R}_{5}$ to obtain the estimated transmit factor matrix ${\boldsymbol{\widehat{C}}_{t}}$ and the receive factor matrix ${{\boldsymbol{\widehat{C}}_{r}}}$ according to (\ref{32})-(\ref{34})
		\State construct the select matrices $\boldsymbol{\Gamma}_{t1}$, $\boldsymbol{\Gamma}_{t2}$, $\boldsymbol{\Gamma}_{r1}$ and $\boldsymbol{\Gamma}_{r2}$ to compute $\boldsymbol{\breve{\Phi}}(\theta_{t}) $ and $\boldsymbol{\breve{\Phi}}(\theta_{r}) $, respectively. And, by using SVD to obtain the estimated $\breve{\theta}_{tk},k=1,2,\cdots,K$ and $\breve{\theta}_{rk},k=1,2,\cdots,K$ according to (\ref{43})-(\ref{50})
		\State estimate $(\boldsymbol{\phi}_{t},\boldsymbol{\gamma}_{t},\boldsymbol{\eta}_{t})$ by using the reconstructed matrix $\boldsymbol{Q}_{t}$ accoding to (\ref{53})-(\ref{57})  
		\State estimate  the receive 3-D parameters $(\boldsymbol{\phi}_{r},\boldsymbol{\gamma}_{r},\boldsymbol{\eta}_{r})$  by using the similar way according to step $\mathbf{8}$ 
		\State  \Return $(\boldsymbol{\theta}_{t},\boldsymbol{\phi}_{t},\boldsymbol{\gamma}_{t},\boldsymbol{\eta}_{t})$ and  $(\boldsymbol{\theta}_{r},\boldsymbol{\phi}_{r},\boldsymbol{\gamma}_{r},\boldsymbol{\eta}_{r})$  
	\end{algorithmic}  
\end{algorithm} 
\subsection{Deterministic Cramer-Rao bound}
According to \cite{WenFShi}, for a bistatic EMVS-MIMO radar with $M$ transmit EMVS and $N$  receive EMVS, the Cramer-Rao bound for the transmit 4-D parameters $(\boldsymbol{\theta}_{t},\boldsymbol{\phi}_{t},\boldsymbol{\gamma}_{t},\boldsymbol{\eta}_{t})$ and the receive 4-D parameters $(\boldsymbol{\theta}_{r},\boldsymbol{\phi}_{r},\boldsymbol{\gamma}_{r},\boldsymbol{\eta}_{r})$   {is}
\begin{equation}
\begin{aligned}
\boldsymbol{CRB}=\frac{\sigma^{2}}{2L}\lbrack real((\boldsymbol{D})^{H} (\boldsymbol{\Pi})^{\perp}_{\boldsymbol{\boldsymbol{C}_{tr}}}  (\boldsymbol{D}) \oplus (\boldsymbol{R}^{T}_{\boldsymbol{ss}}\otimes \boldsymbol{1}_{8\times8} ) )  \rbrack^{-1}
\end{aligned}   
\end{equation}
where $\boldsymbol{C}_{tr}=(\boldsymbol{C}_{t} \odot \boldsymbol{C}_{r})$ {is} the joint transmit-receive matrix of the bistatic EMVS-MIMO radar, $(\boldsymbol{\Pi})^{\perp}_{\boldsymbol{\boldsymbol{C}_{tr}}} =\boldsymbol{I}_{36MN}-\boldsymbol{C}_{tr}\boldsymbol{C}_{tr}^{\dagger}$ denotes the projection matrix of  $\boldsymbol{C}_{tr}$, $\oplus$ denotes the Hadamard product, $\boldsymbol{R}_{\boldsymbol{ss}}$ {represents} the signal covariance matrix, $\boldsymbol{1}_{8\times8}$ {is representative of} the all-one matrix of dimension ${8\times8}$, $\boldsymbol{D}=\begin{bmatrix}
\frac{\partial \boldsymbol{C}_{tr}}{\partial \boldsymbol{\theta}_{t}}&   \frac{\partial \boldsymbol{A}}{\partial \boldsymbol{\phi}_{t}}&
\frac{\partial \boldsymbol{C}_{tr}}{\partial \boldsymbol{\gamma}_{t}}&
\frac{\partial \boldsymbol{C}_{tr}}{\partial \boldsymbol{\eta}_{t})}&
\frac{\partial \boldsymbol{C}_{tr}}{\partial \boldsymbol{\theta}_{r}}&   \frac{\partial \boldsymbol{A}}{\partial \boldsymbol{\phi}_{r}}&
\frac{\partial \boldsymbol{C}_{tr}}{\partial \boldsymbol{\gamma}_{r}}&
\frac{\partial \boldsymbol{C}_{tr}}{\partial \boldsymbol{\eta}_{r}}
\end{bmatrix}$ {is} the joint derivative matrix of the $\boldsymbol{C}_{tr}$ for $(\boldsymbol{\theta}_{t},\boldsymbol{\phi}_{t},\boldsymbol{\gamma}_{t},\boldsymbol{\eta}_{t})$ and  $(\boldsymbol{\theta}_{r},\boldsymbol{\phi}_{r},\boldsymbol{\gamma}_{r},\boldsymbol{\eta}_{r})$.
\subsection{Identified targets}
Assume the maximum number of identified targets is $K$ by using the proposed algorithm in this paper. {The value of $K$ is dependent on the PARAFAC algorithm, owing to the characteristic of PARAFAC decomposition, whose uniqueness, as reported in \cite{GKolda}, can be guaranteed via the following Kruskal's theorem.
}  \\
$\textbf{Theorem 1}$ (Kruskal's theorem) For a three-way tensor model $\mathcal{R}_{5}\in \mathbb{C}^{6\widetilde{M}\times 6\widetilde{N}\times36}$, the PARAFAC decomposition is unique, if
\begin{equation}
\begin{aligned}
\kappa_{{\boldsymbol{\widehat{C}}_{t}}}+ \kappa_{\boldsymbol{\widehat{C}}_{r}}+\kappa_{\boldsymbol{\widehat{Q}}}\geq 2K+2
\end{aligned}   
\end{equation}
where $\kappa_{{\boldsymbol{\widehat{C}}_{t}}}$,$\kappa_{\boldsymbol{\widehat{C}}_{r}}$ and $\kappa_{\boldsymbol{\widehat{Q}}} $ denote the $\kappa$-rank of ${{\boldsymbol{\widehat{C}}_{t}}}$, ${\boldsymbol{\widehat{C}}_{r}}$ and ${\boldsymbol{\widehat{Q}}}$, respectively. The max $\kappa$-rank of ${{\boldsymbol{\widehat{C}}_{t}}}$, ${{\boldsymbol{\widehat{C}}_{r}}}$ and ${\boldsymbol{\widehat{Q}}}$ are   $6\widetilde{M}$,  $6\widetilde{N}$ and  $36$, respectively. Thus, the value of $K$ can be obtained as
\begin{equation}
\begin{aligned}
K=\frac{6\widetilde{M}+6\widetilde{N}+34}{2}
\end{aligned}   \label{60}
\end{equation}
\par {However, the } value of $K$ also {relies} on the rotation invariant relationship in (\ref{47}) and (\ref{48}). The maximum value of $K$ in (\ref{47}) and (\ref{48}) should  satisfy the following constraint
\begin{equation}
\begin{aligned}
\begin{cases}
K\leq 6(\widetilde{M}-1)  \quad in \quad(47)\\
K\leq 6(\widetilde{N}-1)  \quad  in \quad(48)\\
\end{cases}\\ 
\end{aligned}   \label{61}
\end{equation}
\par According to (\ref{60}) and (\ref{61}),  if $\widetilde{M}>\widetilde{N}$, then $K = 6(\widetilde{N}-1)$. If $\widetilde{N}>\widetilde{M}$, then $K = 6(\widetilde{M}-1)$. Thus, the maximum number of identified targets $K$ is
\begin{equation}
\begin{aligned}
K=min\lbrack 6(\widetilde{M}-1),6(\widetilde{N}-1) \rbrack
\end{aligned}   
\end{equation}
\par {Resultantly,}  the maximum number of identified targets is larger than the number of transmit EMVS and receive EMVS. The constructed difference coarrays  in this paper are effective for the underdetermined 2D-DODs and 2D-DOAs estimation in the bistatic EMVS-MIMO radar system .

\subsection{Computational complexity}
\par Table I {compares} the main computational complexity of the proposed algorithm, the ESPRIT algorithm in \cite{Chintagunta},  PM algorithm in \cite{LiuT},  Tensor subspace-based algorithm in \cite{MaoC} and  PARAFAC algorithm in \cite{WenFShi}. {Here,} the CM, SVD and HOSVD denote the covariance matrix calculation, the singular value decomposition and the higher-order singular value decomposition, respectively. For a bistatic EMVS-MIMO radar with $M$ transmit array and $N$ receive array, the computational complexity of the proposed algorithm is higher than that of the PM algotihm and the PARAFAC algorithm, while lower than that of the ESPRIT algorithm and the HOSVD algorithm.

\renewcommand{\arraystretch}{1.5} 
\begin{table*}[!htbp]
	
	\centering
	\fontsize{6.5}{8}\selectfont
	\begin{threeparttable}
		\caption{The computational complexity of different algorithms.}
		\label{tab:performance_comparison}
		\begin{tabular}{cccccc}
			\toprule
			\multirow{2}{*}{Algorithm}&
			\multicolumn{4}{c}{ the complexity of major steps}&  \multirow{2}{*}{ total complexity}\cr
			\cmidrule(lr){2-5} 
			&CM &SVD &HOSVD &multiple iterations\cr
			\midrule
			ESPRIT&$(36MN)^{2}L$&$(36MN)^{3}$&$\times$&$\times$&$\mathcal{O}((36MN)^{2}L+(36MN)^{3} $ \cr
			PM&$(36MN)^{2}L$&$\times$ &$\times$&$\times$&$\mathcal{O}((36MN)^{2}L)$ \cr
			Tensor&$(36MN)^{2}L$&$\times$&$4(36MN)^{3}$&$\times$&$\mathcal{O}((36MN)^{2}L+4(36MN)^{3})$ \cr
			\multirow{2}{*}{PARAFAC}  &\multirow{2}{*}{$\times$}
			&\multirow{2}{*}{$\times$}
			&\multirow{2}{*}{$\times$}
			&$\kappa(3K^{3}+108MNKL+$ & $\mathcal{O}(\kappa(3K^{3}+108MNKL+)$
			\\
			&&&&$3K^{2}(36MN+6NL+6ML))$ &$3K^{2}(36MN+6NL+6ML))$
			\cr
			\multirow{2}{*}{Proposed}  &\multirow{2}{*}{$(36MN)^{2}L$}
			&\multirow{2}{*}{$\times$}
			&\multirow{2}{*}{$\times$}
			&$\kappa(3K^{3}+3888\widetilde{M}\widetilde{N}K+$ & $\mathcal{O}((36MN)^{2}L+\kappa(3K^{3}+3888\widetilde{M}\widetilde{N}K+$
			\\
			&&&&$3K^{2}(36\widetilde{M}\widetilde{N}+216\widetilde{N}+216\widetilde{M}))$ &$3K^{2}(36\widetilde{M}\widetilde{N}+216\widetilde{N}+216\widetilde{M}))$
			\cr
			\bottomrule
		\end{tabular}
	\end{threeparttable}
\end{table*}
\section{Simulation Results}
\par
\begin{figure*}[htbp]
	\centering
	\subfigure[]{
		\includegraphics[width=3in]{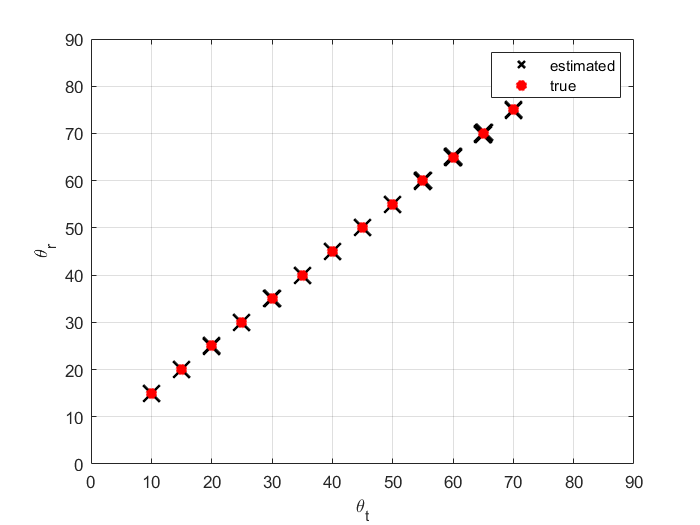}   
	}
	\subfigure[]{
		\includegraphics[width=3in]{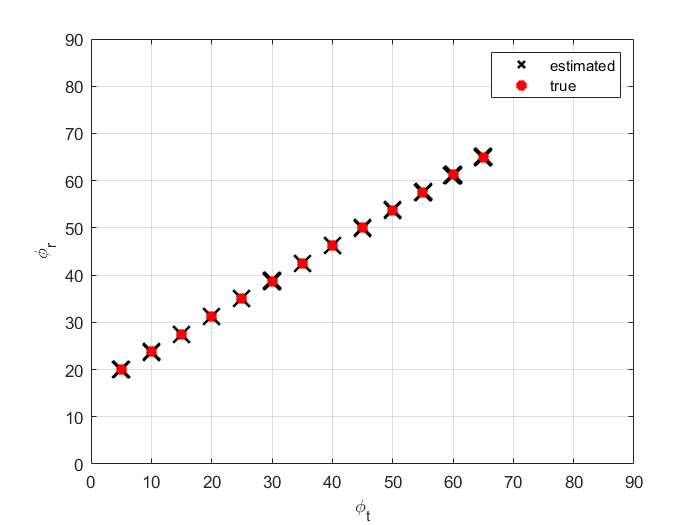}
	}
	
	\subfigure[]{
		\includegraphics[width=3in]{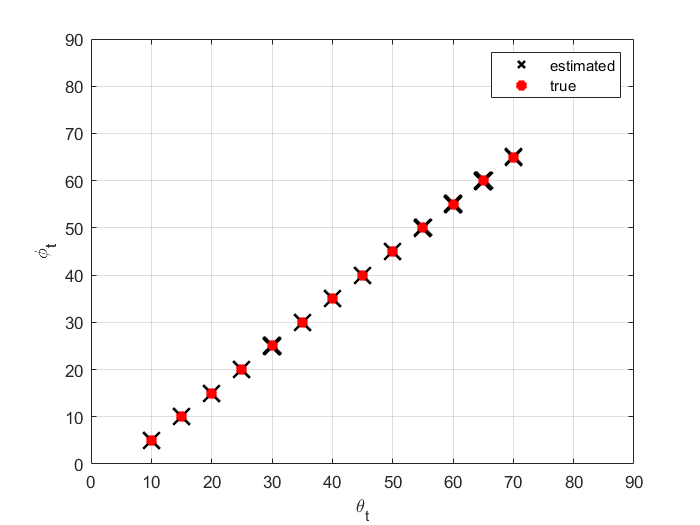}   
	}
	\subfigure[]{
		\includegraphics[width=3in]{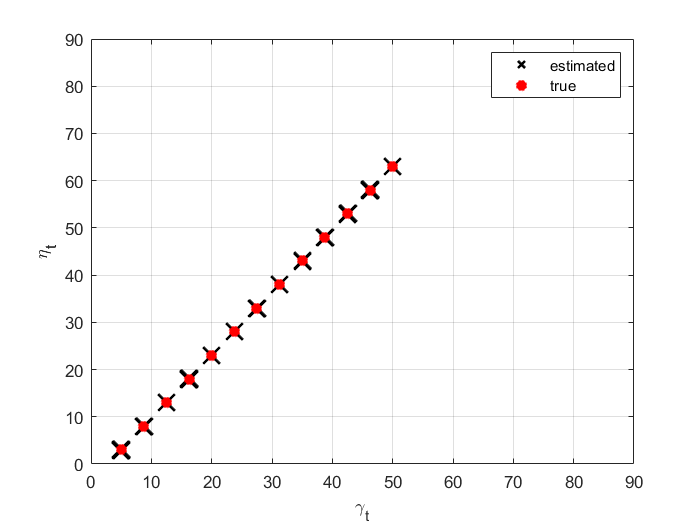}
	}
	
	\subfigure[]{
		\includegraphics[width=3in]{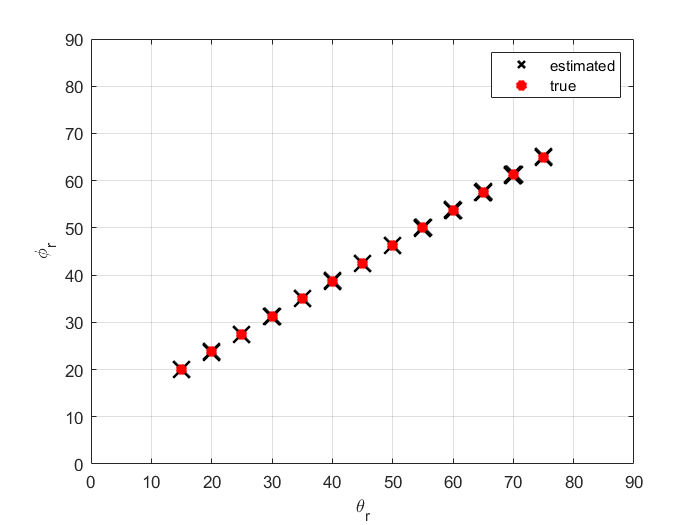}   
	}
	\subfigure[]{
		\includegraphics[width=3in]{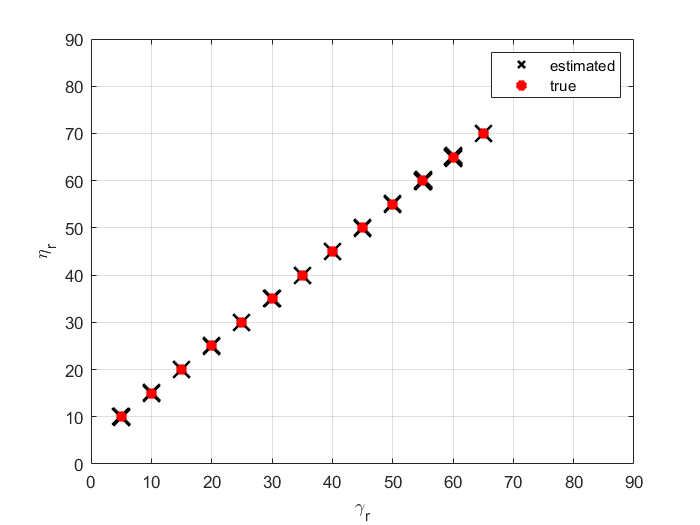}
	}
	\caption{Scatterplot by using the proposed method.
	}
	\label{scatterplot}
\end{figure*}

 In order to evaluate the  estimation performance of angle parameters and polarization parameters in bistatic coprime EMVS-MIMO radar, the following simulation experiments are conducted. $\textbf{Firstly}$,  the underdetermined 2D-DODs and 2D-DOAs estimation performance {is evaluated} by using the proposed algorithm.  Assume the number of transmit EMVS $M$ is $9$ with $(M1,M2)=(3,4)$, the number of receive  EMVS $N$ is $10$ with $(N1,N2)=(3,5)$. The corresponding transmit 4-D parameters and receive 4-D parameters of the impinging targets are listed in Table II. {Additionally,} the SNR is set to $10dB$, the snapshot is $200$.  The noise is {assumed} to be independent zero-mean additive  Gaussian white noise, and the signal and noise are independent of each other. The scatterplot in  Fig. \ref{scatterplot} is obtained via 100 trails. Fig. \ref{scatterplot}(a) and Fig. \ref{scatterplot}(b) show that the underdetermined 2D-DODs and 2D-DOAs are {accurately} estimated. {In the meantime,} Fig. \ref{scatterplot}(c)-Fig. \ref{scatterplot}(f) {demonstrate} that the corresponding transmit 4-D parameters $(\boldsymbol{\theta}_{t},\boldsymbol{\phi}_{t},\boldsymbol{\gamma}_{t},\boldsymbol{\eta}_{t})$ and receive 4-D parameters $(\boldsymbol{\theta}_{r},\boldsymbol{\phi}_{r},\boldsymbol{\gamma}_{r},\boldsymbol{\eta}_{r})$  are  {automatically}  paired {via} the PARAFAC algorithm. Therefore, the effectiveness of  proposed algorithm for the underdetermined 2D-DODs and 2D-DOAs estimation in the bistatic coprime EMVS-MIMO radar has been verified.
\renewcommand{\arraystretch}{1.5} 
\begin{table}[!t]
	\caption{{Parameters of targets}}
	\centering
	\begin{tabular}{cccccc}
		\toprule
		targets &$\boldsymbol{\theta}_{t}/\boldsymbol{\theta}_{r}$ &$\boldsymbol{\phi}_{t}/\boldsymbol{\phi}_{r}$&$\boldsymbol{\gamma}_{t}/\boldsymbol{\gamma}_{r}$&$\boldsymbol{\eta}_{t}/\boldsymbol{\eta}_{r}$  \\
		\midrule
		\multirow{2}{*}{$K=13$}  &{[10:5:70]}
		&{[5:5:65]}
		&{[5:3.75:50]}
		&{[3:5:63]} \\
		&{[15:5:75]}
		&{[20:3.75:65]}
		&{[5:5:65]}
		&{[10:5:70]}\\
		\bottomrule
	\end{tabular}
\end{table}
\par
\begin{figure*}[htbp]
	\centering
	\subfigure[]{
		\includegraphics[width=3in]{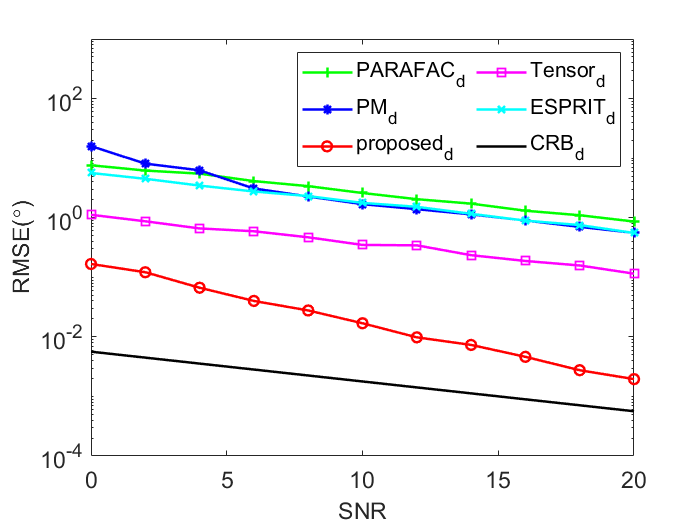}   
	}
	\subfigure[]{
		\includegraphics[width=3in]{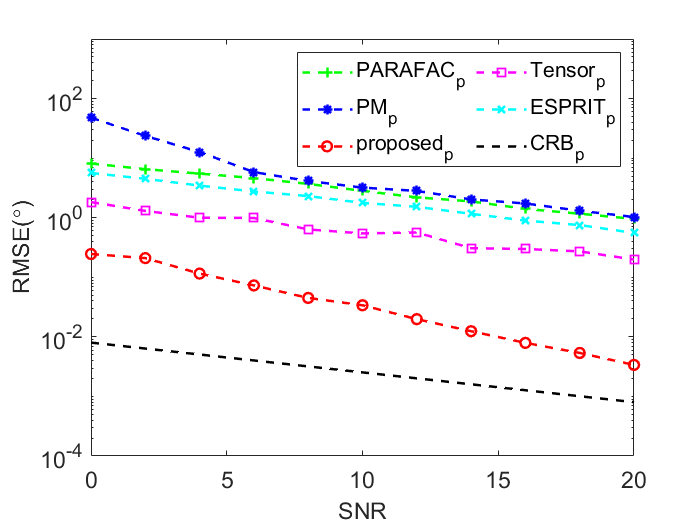}
	}
	
	\caption{RMSE versus SNR.
		(a) angle parameters.   (b) polarization paramters. }
	\label{snr}
\end{figure*}

$\textbf{Secondly}$, the RMSEs performance of different algorithms versus SNR are compared. The average root mean square error (RMSE) is {calculated} as
\begin{equation}
\begin{aligned}
RMSE = \sqrt{\frac{1}{KI}\sum_{i=1}^{I}\|\hat{\boldsymbol{\alpha}}-\boldsymbol{\alpha}\|^{2} }
\end{aligned}   
\end{equation}
where $I=200$ denotes the Monte Carlo trails, $\boldsymbol{\alpha}$ and $\hat{\boldsymbol{\alpha}}$ {represent} the ture angle parameters and estimated angle parameters, respectively. Assume that there are $K=3$ far-field targets with $\boldsymbol{\theta}_{t}=[40^{\circ},20^{\circ},30^{\circ}]$,  $\boldsymbol{\phi}_{t}=[15^{\circ},25^{\circ},35^{\circ}]$, $\boldsymbol{\gamma}_{t}=[10^{\circ},22^{\circ},35^{\circ}]$, $\boldsymbol{\eta}_{t}=[38^{\circ},48^{\circ},56^{\circ}]$, $\boldsymbol{\theta}_{r}=[24^{\circ},38^{\circ},16^{\circ}]$, $\boldsymbol{\phi}_{r}=[21^{\circ},32^{\circ},55^{\circ}]$, $\boldsymbol{\gamma}_{r}=[42^{\circ},33^{\circ},60^{\circ}]$, $\boldsymbol{\eta}_{r}=[17^{\circ},27^{\circ},39^{\circ}]$, respectively. The number of SNR {is increased} from $0dB$ to $20dB$ and the snapshot is set to $200$ for different {SNRs}. The {suffixes} '-d'and '-p' in the legend refer to angle parameter and polarization parameter, respectively. Fig. \ref{snr} {shows} that the proposed algorithm exhibits a better angle and polarization parameters estimation performance compared with the {cutting-edge}  algorithms {on} the conditions of the different SNR.

\par
\begin{figure*}[htbp]
	\centering
	\subfigure[]{
		\includegraphics[width=3in]{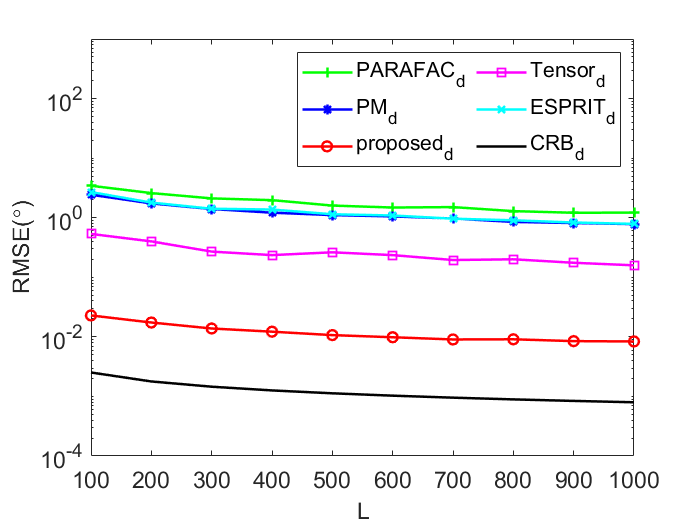}   
	}
	\subfigure[]{
		\includegraphics[width=3in]{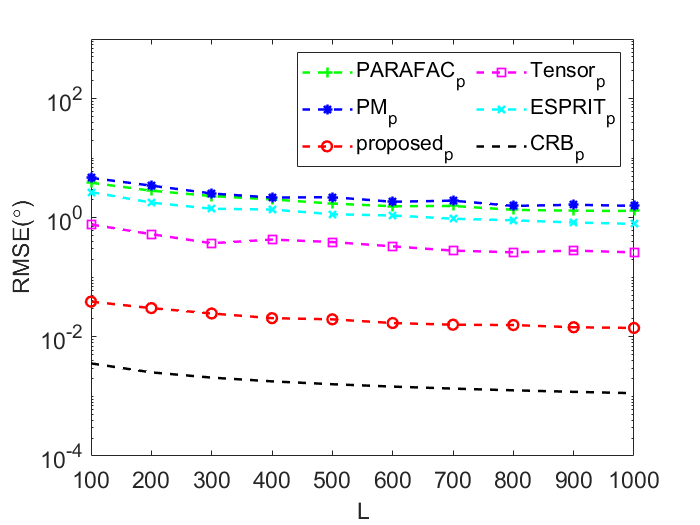}
	}
	
	\caption{RMSE versus snapshot.
		(a) angle parameters.    (b) polarization paramters.  }
	\label{snapshot}
\end{figure*}

$\textbf{Thirdly}$, the RMSEs performance of different algorithms versus the number of snapshots are {assessed}. The snapshots are set to vary from $100$ to $1000$ with a step of $100$ and the SNR is fixed at $10dB$. Other simluation parameters are the same as those in the $\textbf{Secondly}$ experiment.  {As evident from Fig. \ref{snapshot},} the RMSE versus snapshots of the proposed algorithm 
has the smallest estimation error both for the angle parameters and polarization parameters. {As a consequence,} the difference coarray of the copirme EMVS used in this paper promote the  angle parameters and polarization parameters estimation performance.  
\par
\begin{figure*}[htbp]
	\centering
	\subfigure[]{
		\includegraphics[width=3in]{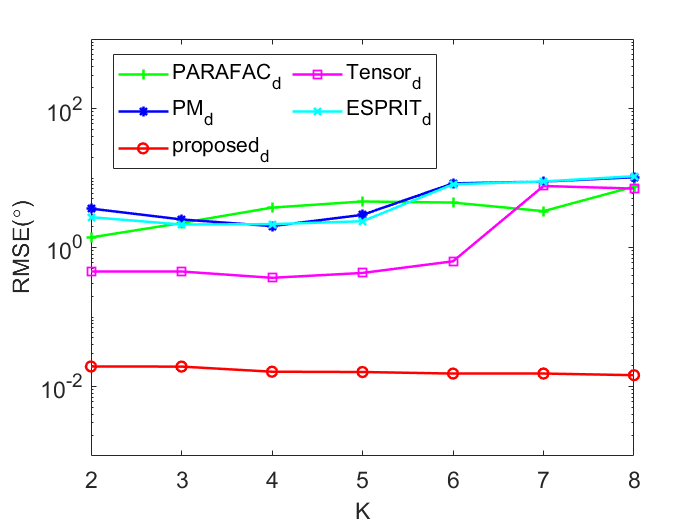}   
	}
	\subfigure[]{
		\includegraphics[width=3in]{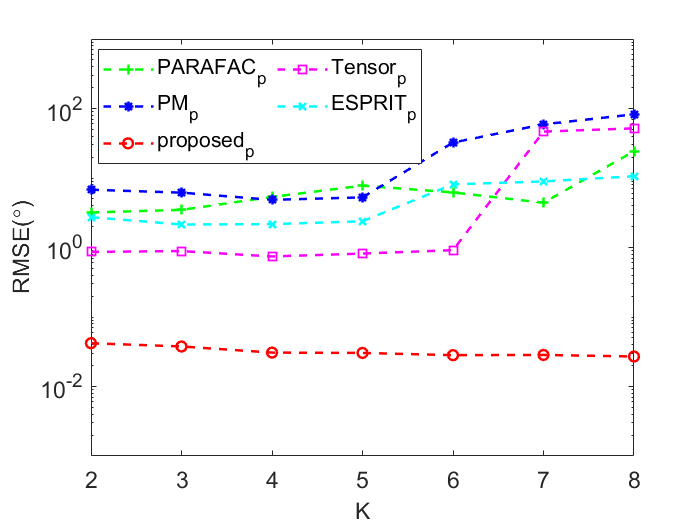}
	}
	
	\caption{RMSE versus $K$.
		(a) angle parameters.    (b) polarization paramters.  }
	\label{RMSEK}
\end{figure*}

$\textbf{Fourthly}$, the RMSEs performance of different algorithms versus the number of impinging sources $K$ {has been compared}. The number of sources vary from $2$ to $8$ and the corresponding transmit 4-D parameter $(\boldsymbol{\theta}_{t},\boldsymbol{\phi}_{t},\boldsymbol{\gamma}_{t},\boldsymbol{\eta}_{t})$ and receive 4-D parameters $(\boldsymbol{\theta}_{r},\boldsymbol{\phi}_{r},\boldsymbol{\gamma}_{r},\boldsymbol{\eta}_{r})$ are choosed from the Table II. The SNR and snapshots are set to $10dB$ and $200$, respectively. {It can be seen from Fig. \ref{RMSEK} that the proposed method is able to maintain a reasonable estimation error for different number of sources, while the RMSEs for the other three algorithms become large with the increase of $K$.} This phenomenon {indicates} that the proposed algorithm is robust to the different number of sources.

\par
\begin{figure*}[htbp]
	\centering
	\subfigure[]{
		\includegraphics[width=3in]{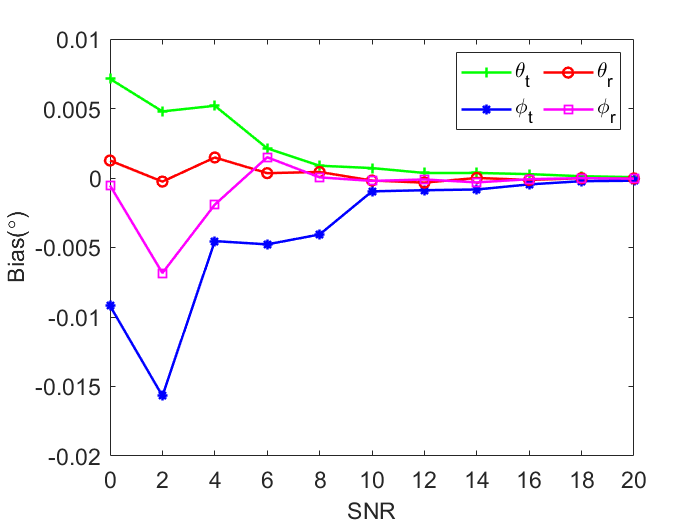}   
	}
	\subfigure[]{
		\includegraphics[width=3in]{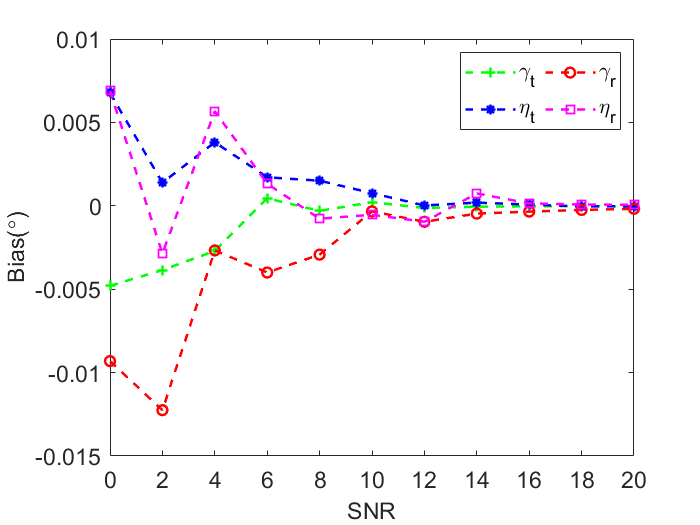}
	}
	
	\subfigure[]{
		\includegraphics[width=3in]{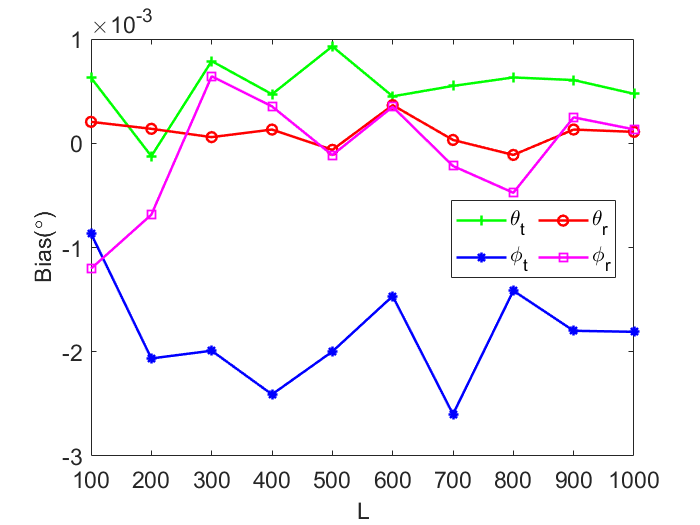}   
	}
	\subfigure[]{
		\includegraphics[width=3in]{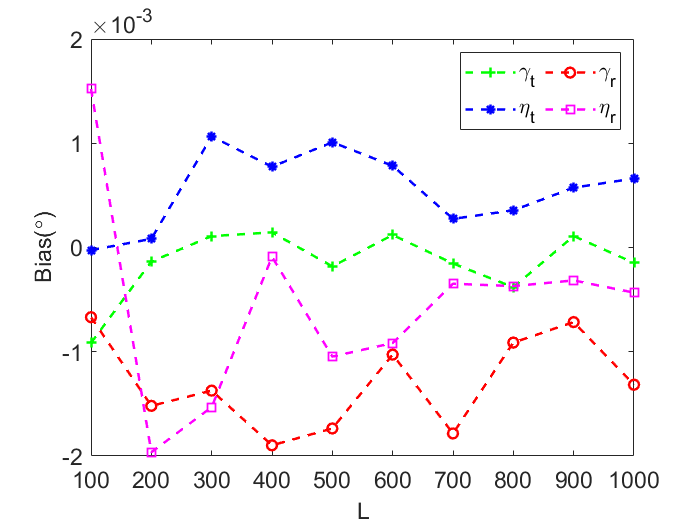}
	}
	\caption{Bias performance of the proposed method for the two closely-located targets.
		(a) angle bias versus SNR . (b) polarization bias versus SNR. (c) angle bias versus snapshot . (d) polarization bias versus snapshot.   }
	\label{anglebias}
\end{figure*}

$\textbf{Finally}$, the estimation performance of the two {closely-}located targets{ has been evaluated} by using the proposed method. Here, the biases between the true angle/polarization parameters and the mean of estimated angle/polarization parameters versus SNR {and} snapshots are examined. The transmit 4-D parameters and receive 4-D parameters of the two closely located targets are set {to}  $\boldsymbol{\theta}_{t}=[22^{\circ},23^{\circ}]$,  $\boldsymbol{\phi}_{t}=[26^{\circ},28^{\circ}]$, $\boldsymbol{\gamma}_{t}=[45^{\circ},55^{\circ}]$, $\boldsymbol{\eta}_{t}=[53^{\circ},63^{\circ}]$, $\boldsymbol{\theta}_{r}=[33^{\circ},35^{\circ}]$, $\boldsymbol{\phi}_{r}=[34^{\circ},36^{\circ}]$, $\boldsymbol{\gamma}_{r}=[20^{\circ},65^{\circ}]$, $\boldsymbol{\eta}_{r}=[28^{\circ},47^{\circ}]$, respectively. {As illustrated in Fig. \ref{anglebias},}  the estimated bias  by  the proposed method is very small with the increase of SNR/snapshots. 

\section{Conclusion}
{A} tensor-based algorithm has been {investigated} to deal with the underdetermined 2D-DOD and 2D-DOA estimation in the bistatic EMVS-MIMO radar system{, and its effectiveness has been well verified via simulations.}
{This algorithm is advantageous to the realization of de-coupling between the spatial response and steering matrix, and it also contributes to the difference coarrays construction of transmit and receive EMVS, which plays an important role to estimate more targets than elements.
}
{More meaningfully,  a variety of newly-designed sparse arrays in recent research can also be used as the transmit/receive EMVS in the bistatic MIMO radar system and this method can pave a path to obtain the corresponding difference coarrays to enhance the estimation performance.
}

\ifCLASSOPTIONcaptionsoff
  \newpage
\fi

\end{document}